\documentclass[final,3p,times,sort&compress,twocolumn]{elsarticle}

\pdfoutput=1
\usepackage{graphicx}

\usepackage[usenames,dvipsnames]{xcolor}  
\usepackage{amssymb,amsfonts,amsmath}

\usepackage{relsize}
\usepackage{placeins}
\usepackage{capt-of}

\newcommand\Tstrut{\rule{0pt}{1.0\normalbaselineskip}}
\newcommand\Bstrut{\rule[-0.4\normalbaselineskip]{0pt}{0pt}}

\def\eps{\varepsilon}

\DeclareMathOperator{\sgn}{sgn}

\newcommand\GR{{G_{\rm R}}}
\newcommand\Grr{{G_{\rm rr}}}

\newcommand\Gqr{{G_{\rm qr}}}
\newcommand\Gpr{{G_{\rm pr}}}
\newcommand\Gqrd{{G_{\rm qrd}}}
\newcommand\Gqrnd{{G_{\rm qrnd}}}
\newcommand\Wqr{{W_{\rm qr}}}
\newcommand\Wpr{{W_{\rm pr}}}
\newcommand\Wrr{{W_{\rm rr}}}

\newcommand\WR{{W_{\rm R}}}

\begin{document} 

\title {Google matrix analysis of 
bi-functional SIGNOR network of protein-protein interactions}

\author[LPT]{Klaus M. Frahm}

\author[LPT]{Dima L.~Shepelyansky}

\address[LPT]{\mbox{Laboratoire de Physique Th\'eorique, IRSAMC, 
Universit\'e de Toulouse, CNRS, UPS, 31062 Toulouse, France}}

\ead[url]{http://www.quantware.ups-tlse.fr/dima}




\begin{abstract}
 Directed protein networks with only 
a few thousand of nodes are rather complex and do not allow to 
extract easily the effective influence of one protein to another
taking into account all indirect pathways via the global network. 
Furthermore, the different types of activation and inhibition actions
between proteins provide a considerable challenge 
in the frame work of network analysis. At the same time these protein 
interactions are of crucial 
importance and at the heart of cellular functioning.
We develop the Google matrix analysis of 
the protein-protein network from the open public database SIGNOR.
The developed approach takes into account the bi-functional activation or 
inhibition nature of interactions between each pair of proteins 
describing it in the frame work 
of Ising-spin matrix transitions. We also apply a recently developed
linear response theory for the Google matrix which highlights a 
pathway of proteins 
whose PageRank probabilities are most sensitive with respect 
to two proteins selected for the analysis. 
This group of proteins is analyzed by 
the reduced Google matrix algorithm which allows to determine the 
effective interactions between them due to direct and indirect pathways 
in the global network. 
We show that the dominating activation or inhibition function of 
each protein can be characterized by its magnetization. 
The results of this Google matrix analysis are presented for three 
examples of selected pairs of proteins. 
The developed methods work rapidly and efficiently
even for networks with several million of nodes and can be applied to various 
biological networks.
\end{abstract}

\maketitle


\section{Introduction}
\label{sec1}

Protein-protein interactions (PPI) are at the heart of information processing
and signaling in cellular functions. It is natural to present
and analyze these PPI by presenting them as a directed network of 
actions between proteins (or network nodes). 
The simplest case of action is activation or inhibition
so that such networks can be considered as bi-functional.
The development of related academic databases of PPS networks
with an open public access is a challenging task with 
various groups working in this direction
(see e.g. 
\cite{pathref1}, \cite{signor}, \cite{pathref2}, 
\cite{pathref3}, \cite{pathref4}).
A typical example is the SIGNOR directed network of PPI links
for about 4000 proteins of mammals and 12000 
bi-functional directed links as reported by  \cite{signor}.

On the scale of the past twenty years, modern society 
has created a variety of complex
communication and social
networks including the World Wide Web (WWW), Facebook, Twitter,
Wikipedia. The size of these networks varies from a several millions
for Wikipedia to billions and more for Facebook and WWW.
The description of generic features of these complex networks 
can be found e.g. in \cite{dorogovtsev}.

An important tool for the analysis of directed networks
is the construction of the Google matrix of Markov transitions
and related PageRank algorithm invented by Brin and Page 
in 1998 for ranking of all WWW sites
(see \cite{brin}, \cite{meyer}). 
This approach has been at the foundations of the Google search engine
used world wide. A variety of applications
of Google matrix analysis to various directed networks
is described by \cite{rmp2015}.

Here we apply recently developed extensions of Google matrix analysis,
which include the REduced GOogle MAtriX (REGOMAX) algorithm 
\cite{regomax} and  the LInear Response algorithm for
GOogle MAtriX (LIRGOMAX) \cite{lirgomax},
to the SIGNOR PPI network.
The efficiency of these algorithms has
been demonstrated for 
Wikipedia networks of politicians \cite{regomax}
and world universities \cite{lirgomax}, \cite{wrwu2017}
and multi product world trade of UN COMTRADE database
\cite{wtn2019}. Thus it is rather natural
to apply these algorithms to PPI networks
which have a typical size being significantly smaller
than Wikipedia and WWW.

From a physical view-point the LIRGOMAX 
approach corresponds to a small probability pumping  
at a certain network node (or group of nodes) and
absorbing probability at another specific node (or group
of nodes). This algorithm allows first to determine the
most sensitive group of nodes involved in this 
pumping-absorption process tracing a pathway connecting two selected proteins. 
In a second stage one can then apply the REGOMAX algorithm and obtain 
an effective reduced Google matrix, and in particular effective interactions, 
for the found subset of most sensitive nodes. 
These interactions are due to either direct or indirect pathways 
in the global huge network in which is embedded the selected relatively small
subset of nodes. 

The REGOMAX and LIRGOMAX algorithms
originate from the scattering theory of nuclear and mesoscopic physics, 
field of quantum chaos and linear response theory of
electron transport \cite{regomax}, \cite{lirgomax}.

We point out that the analysis of the SIGNOR PPI network
already found biological applications reported
by \cite{signorcell}, \cite{naturebio}, \cite{petre}, \cite{niko}.
The detailed review of various applications of the PPI  
signaling networks is given by \cite{beltrao}.
However, the Google matrix analysis has not been used in these
studies.

The challenging feature of PPI networks is the bi-functionality of 
directed links which produce activation or inhibition actions. 
While in our previous analysis of SIGNOR network by \cite{zinovyev}
this feature was ignored, here we apply the Ising-PageRank
approach developed in \cite{isingnet}
for opinion formation modeling. In this Ising-type approach
the number of nodes in the PPI network is doubled, with a $(+)$ or 
$(-)$ attribute for each protein, 
and the links between doubled nodes
are described by $2 \times 2$ matrices 
corresponding to activation or inhibition actions.

In this work we apply the LIRGOMAX and REGOMAX
algorithm to the bi-functional PPI network of SIGNOR.
We show that this approach allows to determine the 
effective sensitivity with direct and indirect interactions
between a selected pair of proteins. As particular examples we will choose 
three protein pairs implicating the 
{\it Epidermal growth factor receptor (EGFR)} 
which is considered to play an important role in the context of lung cancer 
(see e.g. \cite{egfr}, \cite{krasnoyarsk}).

The interest to apply computer science methods,
such as the PageRank algorithm, to PPI networks is growing
(see e.g. the recent review \cite{cowen}) and we hope that the 
Google matrix algorithms
described in this work will attract the interest of biologists
working with PPI networks. In addition to the methods described in \cite{cowen}
these algorithms allow to take into account the bi-functional nature
of PPI network links and focus the investigation on a specific group
of proteins taking into account all their direct and indirect interactions
via the global network.

The paper is constructed as follows:
in Section 2 we describe the construction of Google matrix 
from links between proteins and related 
LIRGOMAX and REGOMAX algorithms, in Section 3 we characterize 
data sets and the Ising-PPI-network
for bi-functional interactions between proteins,
results are presented in Section 4
and the conclusion is given in Section 5.
Appendix provides additional matrix data 
and executable code for the described algorithms
for the SIGNOR Ising-PPI-network.

The Google matrix data and executive code of described algorithms 
are available at http://www.quantware.ups-tlse.fr/QWLIB/google4signornet/ .

\section{Methods of Google matrix analysis}

\subsection{Google matrix construction}

The Google matrix $G$ of $N$ nodes (proteins or proteins with $(+)/(-)$ 
attribute) is constructed from 
the adjacency matrix $A_{ij}$ with element $1$ 
if node $j$ 
points to  node $i$ and zero otherwise. 
The matrix $G$ has the standard form 
$G_{ij} = \alpha S_{ij} + (1-\alpha) / N$ 
(see \cite{brin}, \cite{meyer}, \cite{rmp2015}),
where $S$ is the matrix of Markov transitions with elements  
$S_{ij}=A_{ij}/k_{out}(j)$ and $k_{out}(j)=\sum_{i=1}^{N}A_{ij}\neq 0$ 
being the  out-degree of node $j$ 
(number of outgoing links);  $S_{ij}=1/N$ if $j$ 
has no outgoing links (dangling node). 
The parameter $0< \alpha <1$ is known as the damping factor
with the usual value $\alpha=0.85$ \cite{meyer}
which we use here.
For the range $0.5 \leq \alpha \leq 0.95$
the results are not sensitive to $\alpha$ \cite{meyer}, \cite{rmp2015}. 
A useful view on this $G$ matrix is given by the concept of  
a random surfer, moving with probability $\alpha$ from one node to another 
via one of the available directed links 
or with a jump probability $(1-\alpha)$ to any node.

The right PageRank eigenvector
of $G$ is the solution of the equation $G P = \lambda P$
for the leading unit eigenvalue $\lambda=1$ \cite{meyer}. The PageRank 
$P(j)$ values represent positive probabilities to find a random surfer 
on a node $j$ ($\sum_j P(j)=1$). 
All nodes can be ordered by decreasing probability $P$ 
numbered by  PageRank index $K=1,2,...N$ with a maximal probability at $K=1$
and minimal at $K=N$. The numerical computation 
of $P(j)$ is done efficiently with the PageRank iteration
algorithm described by \cite{meyer}. The idea of this algorithm is simply 
to start with some initial, sum normalized, vector $P^{(0)}$ of positive 
entries, e.g. being $1/N$ for simplicity, and then to iterate 
$P^{(n+1)}=G\,P^{(n)}$ which typically converges after $n=150-200$ iterations 
(for $\alpha=0.85$). 

It is also useful to consider
the original network with inverted direction of links.
After inversion the Google matrix $G^*$ is constructed via
the same procedure with $G^* P^*= P^*$. The matrix $G^*$  
has its own PageRank vector
$P^*(j)$ called CheiRank \cite{cheirank}, \cite{rmp2015}.
Its values give probabilities to find a random surfer of a given node
and they can be again ordered
in a decreasing order with CheiRank index $K^*$
with highest  $P^*$ at $K^*=1$ and smallest at $K^*=N$.
On average, the high values of $P$ ($P^*$) correspond to nodes
with many ingoing (outgoing) links \cite{meyer}, \cite{rmp2015}.

\subsection{Reduced Google matrix (REGOMAX) algorithm}

The REGOMAX algorithm is described in detail 
by  \cite{regomax,zinovyev,wrwu2017}. It allows to 
compute efficiently a ``reduced Google matrix'' 
$\GR$ of size $N_r \times N_r$ 
that captures
the full contributions of direct and indirect pathways appearing 
in the full Google matrix $G$ between $N_r$ nodes of interest
selected from a huge global network with $N \gg N_r$ nodes. 
For these $N_r$ nodes 
their PageRank probabilities are the same 
as for the global network with $N$ nodes, 
up to a constant multiplicative factor taking into account that 
the sum of PageRank probabilities over $N_r$
nodes is unity.   
The computation of $\GR$ determines 
a decomposition of $\GR$ into matrix components that clearly distinguish 
direct from indirect interactions: 
$\GR = \Grr + \Gpr + \Gqr$ \cite{regomax}.
Here $\Grr$ is given by the direct links between the selected 
$N_r$ nodes in the global $G$ matrix with $N$ nodes. We note that 
 $\Gpr$ is rather close to 
the matrix in which each column is approximately proportional to 
the PageRank vector $P_r$, satisfying the condition that the 
PageRank probabilities of $\GR$ are 
the same as for $G$ (up to a constant multiplier due to normalization).
Hence, in contrast to $\Gqr$, $\Gpr$ doesn't give much new 
information about direct and indirect links between selected nodes.

The most interesting role is played by $\Gqr$, which takes 
into account all indirect links between
selected nodes happening due to multiple pathways via 
the global network of nodes $N$ (see~\cite{regomax}).
The matrix  $\Gqr = \Gqrd + \Gqrnd$ has diagonal ($\Gqrd$)
and non-diagonal ($\Gqrnd$) parts with $\Gqrnd$ 
describing indirect interactions between selected nodes.
The exact formulas for all three components of $\GR$ are given 
in \cite{regomax}. It is also useful 
to compute the weights $\WR$, $\Wpr$,
$\Wrr$, $\Wqr$
of $\GR$ and its 3 matrix components $\Gpr$, $\Grr$, $\Gqr$
given by the sum of all its elements divided by the matrix size $N_r$.
Due to the column sum normalization of $\GR$ we obviously have 
$\WR=\Wrr+\Wpr+\Wqr=1$. 

We note that the matrix elements of $\Gqr$
may have negative values (only the full reduced matrix
$\GR$ should have positive elements; 
$\Grr$ also has only positive matrix elements)
but these negative values are found to be small
for the Ising-PPI-networks
and do not play a significant role.
A similar situation for Wikipedia networks
is discussed by \cite{regomax}, \cite{lirgomax}.

\subsection{LIRGOMAX algorithm}

The detained description of the LIRGOMAX algorithm
is given by \cite{lirgomax}. It performs an infinitely weak
$\varepsilon$-probability injection (pumping) at one node 
(a protein or a protein with $(+)/(-)$ attribute)
and absorption at another node of interest. This process is described 
by the modified PageRank iteration $P^{(n+1)}=G\,F(\eps,P^{(n)})$ 
where the vector valued function $F(\eps,P)$ has the 
components $P(i)+\eps$ for $i$ being the index of the injection/pumping node, 
$P(j)-\eps$ for $j$ being the index of the absorption node 
and simply $P(k)$ for all other nodes $k$. In this way the vector $F(\eps,P)$ 
is also sum normalized if $P$ is sum normalized and obviously $F(0,P)=P$ is 
the identity operation. 
In \cite{lirgomax} a more general version of $F(\eps,P)$ was considered 
with potentially different prefactors for the $\eps$ contributions, 
injection/absorption at possibly more than two nodes and an 
additional renormalization factor to restore the sum normalization (which 
is automatic in the simple version). However, for 
the applications in this work the above given simple version of $F(\eps,P)$ 
is sufficient. 

In principle one can solve iteratively the above 
modified PageRank iteration formula which converges at the same rate as 
the usual PageRank iteration algorithm and provides a modified 
$\eps$-depending PageRank $P(\eps)$. Then one can compute 
the linear response vector 
$P_1=dP(\eps)/d\eps|_{\eps=0}=\lim_{\eps\to 0} [P(\eps)-P_0]/\eps$
 where $P_0$ is the PageRank 
obtained for $\eps=0$. However the naive direct evaluation of this limit 
is numerically not stable in the limit $\eps\to 0$. 
Fortunately as shown by \cite{lirgomax} it is possible 
to compute $P_1$ directly 
in an accurate and efficient way by solving the inhomogeneous PageRank 
equation 
\begin{equation}
\label{eq_PGinhom}
P_1=G\,P_1+V_0\quad,\quad V_0=G\,W_0
\end{equation}
where the vector $W_0$ has only two non-zero components for the two 
particular injection or absorption nodes $W_0(i)=1$ or $W_0(j)=-1$ 
respectively. Therefore a more explicit expression for the vector 
$V_0$ appearing in (\ref{eq_PGinhom}) is 
$V_0(k)=G_{ki}-G_{kj}$ (for all nodes $k$). We mention that 
the three vectors $P_1$, $V_0$ and $W_0$ are orthogonal to the vector 
$E^T=(1,\ldots,1)$ composed of unit entries, i.e. 
$\sum_k P_1(k)=\sum_k W_0(k)=\sum_k V_0(k)=0$. Furthermore, all of these 
vectors, especially $P_1$ 
have {\em real} positive or negative entries (note that 
in general eigenvectors of a non-symmetric real matrix may be complex).

A formal solution of the inhomogeneous PageRank equation is: 
$P_1=\sum_{n=0}^\infty G^n\,V_0=({\bf 1}-G)^{-1}\,V_0$ 
which is well defined since $V_0$, when expanded in the basis 
of (generalized) eigenvectors of $G$, does NOT have a contribution 
of $P_0$ (the only eigenvector of $G$ with eigenvalue $1$) such that 
the singularity of the matrix inverse does not constitute a problem. 
Of course numerically, we compute $P_1$ in a different way, as described 
by \cite{lirgomax} one can iterate the equation 
$P_1^{(n+1)}=G\,P_1^{(n)}+V_0$ with $P_1^{(0)}=0$ which converges with the 
same rate as the usual PageRank iteration. 

We note that a propagator somewhat similar to the above expression 
$P_1=({\bf 1}-G)^{-1}\,V_0$,
namely $\tilde{P} = ({\bf 1} - \gamma G)^{-1}\,V_{init}$,
has been used in \cite{physrevnet} as the ImpactRank of specific nodes
related to an initial probability localized on a certain initial node
described by the initial vector $V_{init}$. However, this ImpactRank 
used $\gamma<1$
so that there was no singularity in denominator
and also $\tilde{P}$ represented a certain 
stationary probability distribution while $P_1$ represents a deviation
from the stationary distribution of PageRank probability $P$.
In fact the propagator discussed in \cite{cowen}
corresponds to the ImpactRank case \cite{physrevnet} 
with $\gamma <1$ thus being qualitatively different
from the LIRGOMAX propagator considered here.

In a similar as the PageRank $P_0$ is characterized by the index $K$ we 
introduce the index $K_L$ by ordering $|P_1|$ such that $K_L=1$ corresponds 
to the node with largest value of $|P_1|$ and $K_L=N$ to the node with 
smallest value of $|P_1|$. 
Once $P_1$ is computed for the pair of chosen injection/absorption nodes 
we determine the 20 top nodes with strongest negative values of $P_1$ and 
further 20 top nodes with strongest positive values of $P_1$ which 
constitute a subset of 40 nodes which are the most significant
nodes participating in the pathway between the pumping node $i$ and 
absorbing node $j$. We also require that these two particular nodes $i$ and 
$j$ belong to this subset. If this is not automatically the case we replace 
the node at total position 20 (position 20 for strongest negative values 
of $P_1$) with the absorption node $j$ and/or the node at total position 40 
(position 20 for strongest positive values of $P_1$) with the injection node 
$i$. This situation happens once for the absorption node of the third example 
below which has a very low ranking position $K_L\approx 2000$ for $|P_1|$. 

In general from a physical/biological point of view we indeed expect that 
the two particular injection/absorption nodes belong automatically to the 
selected subset of most sensitive nodes. However, there is no simple 
or general mathematical argument for this. 

Using this subset of top nodes in the $K_L$ ranking 
we then apply the REGOMAX algorithm 
to compute the reduced Google matrix and its components and in 
particular we determine the effective direct and indirect interactions of this 
reduced network. 
The advantage of the application of LIRGOMAX at the initial stage 
is that it provides an automatic and more rigorous 
procedure to determine an interesting 
subset of protein nodes related to the pumping between nodes $i$ and $j$ 
instead of using an arbitrary heuristic choice for such a subset. 

\section{Data sets and Ising-PPI-network construction}

We use the open public SIGNOR PPI network \cite{signor}
(April 2019 release for human, mouse and rat).
This network contains $N=4341$ nodes (proteins)
and $N_\ell = 12547$ directed hyperlinks between nodes.
Each protein (node) is described by their name and identifier.

A new interesting feature of this PPI directed network is that its hyperlinks
have activation and inhibition actions. 
For some links the functionality is unclear and 
then they are considered to be neutral. 
This feature rises an interesting mathematical
challenge for the Google matrix description
of such bi-functional networks.
To meet this challenge we use the Ising-PageRank approach
developed by \cite{isingnet} for a model of opinion formation
on social networks. In this approach each node is doubled getting
two components marked by $(+)$ and $(-)$. 
The activation links point to the $(+)$ components
and inhibition links point to the $(-)$ components. Such transitions
between doubled nodes are described by $2\times 2$ block matrices 
$\sigma_+$ ($\sigma_-$) matrices with entries 1 (0) in the first row 
and 0 (1) in the second row 
as for Ising spin-1/2 (see details described in Appendix).
A neutral transition is described by $2\times2$ matrix $\sigma_0$
with all elements being $1/2$. 
Thus for this Ising-network (doubled-size network) 
we have doubled number of node $N = 8682$
and the total number of hyperlinks being $N_\ell = 27266$;
among them there are $N_{act} = 14944$ activation links,
$N_{inh} = 7978$ inhibition links and
$N_{neut} = 4344$ neutral links ($N_\ell = N_{act} + N_{inh} + N_{neut}$).
From this weighted Ising-PPI-network with $N_\ell = 27266$
nodes we construct the Google matrix 
following the standard rules described by \cite{meyer}, \cite{rmp2015} 
and also given above.

Below we apply the Google matrix analysis
taking into account the bi-functionality PPI 
and illustrate the efficiency
of the LIRGOMAX and REGOMAX algorithms
for the SIGNOR Ising-PPI-network.

The details of Ising-PPI-network construction,
its main statistical properties 
and an executable code for the described algorithms
are provided in Appendix
and in \cite{ourwebpage}.
Below we discuss the results obtained with 
the LIRGOMAX and REGOMAX algorithms
for three examples of specific pathways between two specific proteins.

\section{Results}

Here we present results obtained with LIRGOMAX and
REGOMAX algorithms for 
pathways between several pairs of selected proteins.

\subsection{Case of  pathway EGFR - JAK2 proteins}

\begin{table}
\begin{center}
{
\relsize{-1}
\caption{Top 20 nodes of strongest negative values of 
$P_1$ (index number $i=1,\ldots,20$) and top 20 nodes of strongest 
positive values of $P_1$ (index number $i=21,\ldots,40$) 
with $P_1$ being created as the linear response of the PageRank of 
the Ising-PPI-network
with injection (or pumping)  at {\it  EGFR P00533 (+)} 
and absorption at {\it JAK2 O60674 (-)}; $K_L$ is the ranking index 
obtained by ordering $|P_1|$ and $K$ is the usual PageRank index
obtained by ordering the PageRank $P_0$. 
}
\label{table1}
\begin{tabular}{rrrl}
\hline
$i$ & $K_L$ & $K$ &Node name \Tstrut\Bstrut\\
\hline\Tstrut
1 & 1 & 30 & JAK2	O60674 (+) \\
2 & 4 & 470 & JAK2	O60674 (-) \\
3 & 5 & 554 & IFNGR2/INFGR1	SIGNOR-C142 (+) \\
4 & 6 & 354 & ARHGEF1	Q92888 (+) \\
5 & 7 & 631 & APOA1	P02647 (+) \\
6 & 8 & 956 & CSF2RA/CSF2RB	SIGNOR-C212 (+) \\
7 & 9 & 57 & STAT1	P42224 (+) \\
8 & 10 & 204 & MAP3K5	Q99683 (-) \\
9 & 12 & 1008 & STAT4	Q14765 (+) \\
10 & 13 & 825 & CCR2	P41597 (+) \\
11 & 14 & 2377 & PRMT5	O14744 (-) \\
12 & 15 & 2378 & STAM	Q92783 (+) \\
13 & 16 & 1482 & EPOR	P19235 (+) \\
14 & 17 & 1117 & CSF2RA	P15509 (+) \\
15 & 18 & 959 & ITGAL	P20701 (+) \\
16 & 19 & 1968 & CTLA4	P16410 (-) \\
17 & 20 & 2058 & STAP2	Q9UGK3 (+) \\
18 & 21 & 2024 & ITGB2	P05107 (+) \\
19 & 22 & 532 & EZH2	Q15910 (-) \\
\Bstrut 20 & 23 & 1196 & GTF2I	P78347 (+) \\
\hline\Tstrut
21 & 2 & 29 & GRB2	P62993 (+) \\
22 & 3 & 172 & FES	P07332 (+) \\
23 & 11 & 90 & EGFR	P00533 (+) \\
24 & 27 & 3 & PIK3CD	O00329 (-) \\
25 & 30 & 126 & CBL	P22681 (+) \\
26 & 31 & 136 & EGFR	P00533 (-) \\
27 & 32 & 648 & EZR	P15311 (+) \\
28 & 36 & 38 & PTK2	Q05397 (+) \\
29 & 37 & 456 & GAB1	Q13480 (+) \\
30 & 39 & 424 & BCR	P11274 (-) \\
31 & 42 & 124 & PIK3R1	P27986 (+) \\
32 & 43 & 26 & PLCG1	P19174 (+) \\
33 & 44 & 58 & SHC1	P29353 (+) \\
34 & 45 & 88 & ESR1	P03372 (+) \\
35 & 46 & 2398 & VAV2	P52735 (+) \\
36 & 47 & 746 & SHC3	Q92529 (+) \\
37 & 48 & 291 & ERBB2	P04626 (+) \\
38 & 49 & 888 & ERBB3	P21860 (+) \\
39 & 51 & 1109 & NCK1	P16333 (+) \\
\Bstrut 40 & 52 & 1531 & CRK	P46108 (-) \\
\hline
\end{tabular}
}
\end{center}
\end{table}

As a first example we choose the node {\it EGFR P00533 (+)} for 
injection (pumping) and {\it JAK2 O60674 (-)} for 
absorption. It is known that
mutations affecting the protein EGFR expression or activity
could result in lung cancer (see e.g. \cite{egfr}; 
\cite{krasnoyarsk}). This protein
interacts with the protein JAK2 
whose mutations have been implicated in various types of cancer.
We argue that the injection (pumping) at
{\it EGFR P00533 (+)} and absorption at {\it JAK2 O60674 (-)}
should involve certain variations of the 
PageRank probability, represented by $P_1$,
showing interactions between various proteins
actively participating in the pathway from
{\it EGFR P00533 (+)} to {\it JAK2 O60674 (-)}.
The pumping process can be viewed as a result of
disease development and absorption as a certain mutation
of this disease into another one.

\begin{figure}[h!]
\centerline{\includegraphics[width=0.48\textwidth]{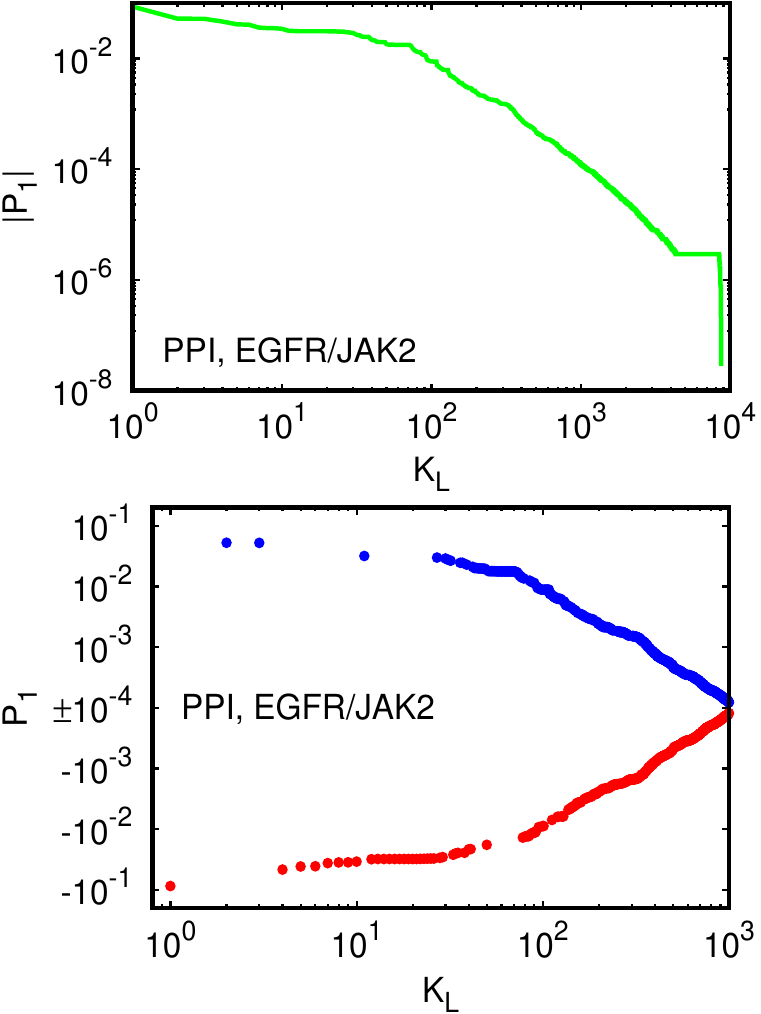}}
\caption{Linear response vector $P_1$ of PageRank for the Ising-PPI-network
with injection (or pumping) at {\it EGFR P00533 (+)} 
and absorption at {\it  JAK2 O60674 (-)}. Here $K_L$ is the ranking index 
obtained by ordering $|P_1|$ from maximal value at $K_L=1$
down to minimal value. Top panel shows $|P_1|$ versus $K_L$ in 
a double logarithmic representation for all $N$ nodes.
Bottom panel shows a zoom of $P_1$ versus $K_L$ for $K_L \le 10^3$ 
in a double logarithmic representation with sign; 
blue data points correspond 
to $P_1>0$ and red data points to $P_1<0$. 
}
\label{fig1}
\end{figure}

The global PageRank indices of these two nodes are
$K= 90$ (PageRank probability $P(90)= 0.0009633$ ) for  {\it EGFR P00533 (+)} and
$K=470$ (PageRank probability $P(470)= 0.0003444$) for {\it JAK2 O60674 (-)}.
As described above in the LIRGOMAX computations we choose the vector 
in $V_0$ which appears in the inhomogeneous PageRank equation 
(\ref{eq_PGinhom}) as $V_0=G\,W_0$ 
with $W_0(K=90)=+1$, $W_0(K=470)=-1$ and $W_0(K)=0$ for all other 
values of the Kindex $K$. 
We remind that both $W_0$ and $V_0$ are orthogonal to the left 
leading eigenvector $E^T=(1,\ldots,1)$ of $G$ according to the 
general description of the LIRGOMAX algorithm given above and 
in \cite{lirgomax}. 

For comparison we let us note that the top 4 PageRank nodes are
$K=1$ ($P(1)=0.003041$) for {\it CASP3	P42574 (+)},
$K=2$ ($P(2)=0.002821$) for {\it NOTCH1	P46531 (+)},
$K=3$ ($P(3)=0.002433$) for {\it PIK3CD	O00329 (-)},
$K=4$ ($P(4)=0.002413$) for $\;\;\;\;\;\;$ {\it CTNNB1	P35222 (-)}
(other values/data are available at \cite{ourwebpage}).

Similar to the two Wikipedia examples analyzed by \cite{lirgomax}  
the LIRGOMAX algorithm selects the
proteins mostly affected by injection/absorption process
with 20 most positive (EGFR block) and 20 most negative 
(JAK2 block) values of $P_1$ shown in Table~\ref{table1}.
Here the pumped protein {\it EGFR P00533 (+)} is on the third position
in its block of positive $P_1$ values ($i=23$) and with $K_L =11$ 
(where $K_L$ is the ranking index obtained by ordering the components 
of $|P_1|$) while the protein with absorption {\it JAK2 O60674 (-)}
has the second position in its block of negative $P_1$ values ($i=2$)
with $K_L=4$. Thus these two nodes are not at the first positions in 
their respective blocks but still they are placed at very high positions.

\begin{figure}[h!]
\centerline{\includegraphics[width=0.48\textwidth]{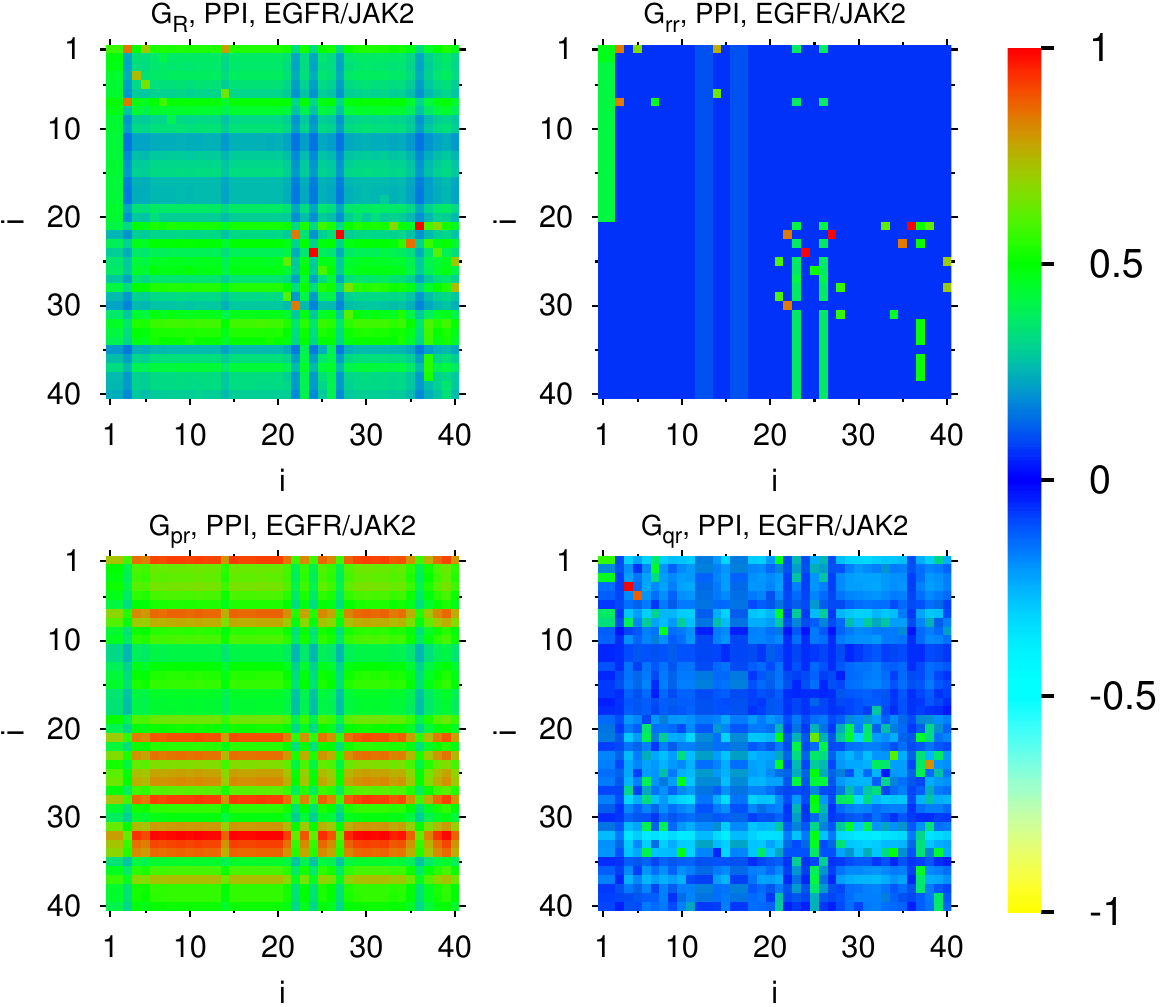}}
\caption{Reduced Google matrix components $\GR$, $\Gpr$, 
$\Grr$ and $\Gqr$ for Ising-PPI-network and the 
subgroup of nodes given in Table~\ref{table1} corresponding to 
injection at {\it EGFR P00533 (+)} 
and absorption at {\it JAK2 O60674 (-)} (see text for explanations). The 
axis labels correspond to the index  $i$ used in Table~\ref{table1}.
The relative weights of these components are $\Wpr=0.761$, $\Wrr=0.220$, 
 and $\Wqr=0.019$. 
The values of the color bar correspond to 
$\sgn(g)(|g|/\max|g|)^{1/4}$ where $g$ is the shown 
matrix element value. The exponent $1/4$ amplifies 
small values of $g$ for a better visibility. 
}
\label{fig2}
\end{figure}

The dependence of $|P_1|$ of the index $K_L$ is shown 
in the top panel of Figure~\ref{fig1}. The decay of 
$|P_1|$ is relatively slow for $K_L \leq 40$
followed by a more rapid drop for $K_L > 40$.
The bottom panel shows the dependence of 
positive (blue) and negative (red)
values of $P_1$ on $K_L$. We note that the top absolute values 
$|P_1|$ for blue and red components have 
comparable values being of the order of $|P_1| \sim 0.1$
for approximately $K_L \leq 40$. However, in this range
the number of positive (blue)
values of $P_1$ is significantly smaller
compared to the number of negative (red) values of $P_1$.
This point can also be seen from the column of $K_L$ values 
in Table~\ref{table1}. 
Another feature visible from Table~\ref{table1}
is that the number of proteins with negative component $(-)$
is significantly smaller than those with a positive component $(+)$
($5$ for $1 \leq i \leq 20$ and $4$ for  
$21 \leq i \leq 40$. We return to 
the properties of positive and negative components a bit later.

\begin{figure}[h!]
\centerline{\includegraphics[width=0.48\textwidth]{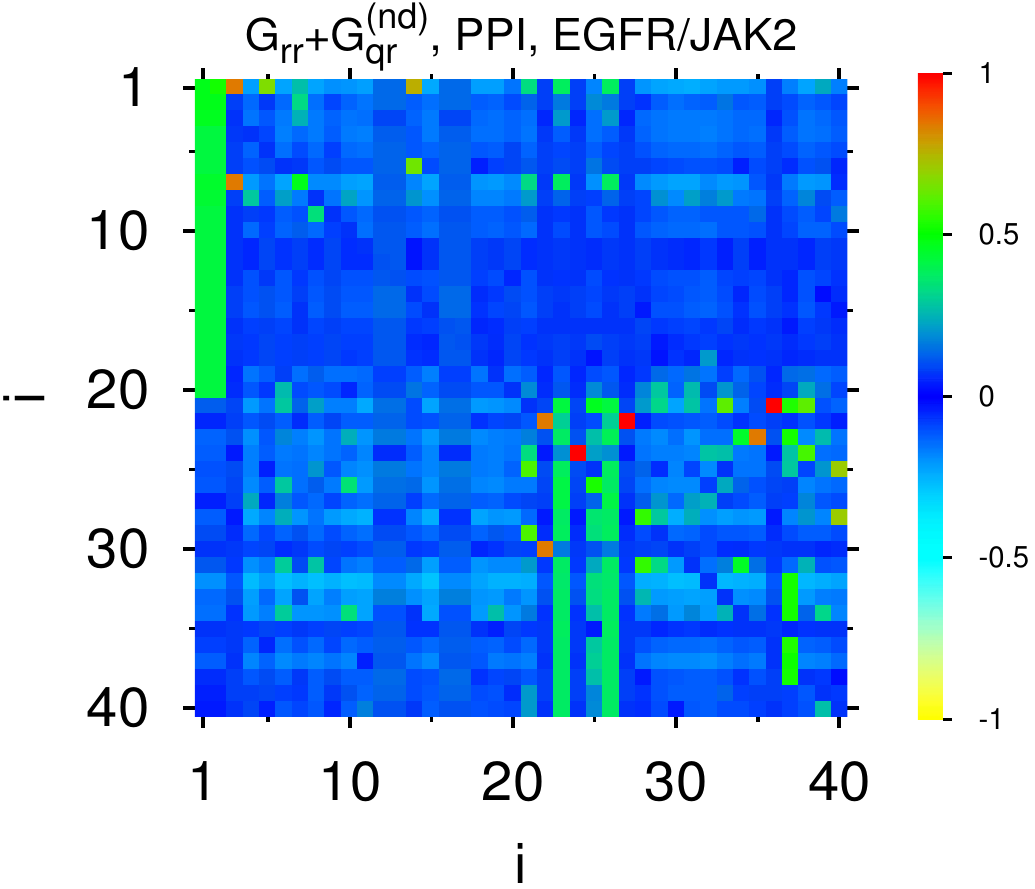}}
\caption{Same as in Fig.~\ref{fig2} but for the matrix $\Grr+\Gqrnd$, 
where $\Gqrnd$ is obtained from $\Gqr$ by putting its diagonal 
elements at zero; the weight of these two components is 
$W_{rr+qrnd}=0.227$.
}
\label{fig3}
\end{figure}

After the selection of most significant 40 nodes of the pathway
between the two injection/absorption proteins (see Table~\ref{table1})
we apply the REGOMAX algorithm
which determines all matrix elements of Markov transitions
between these 40 nodes including all direct and indirect pathways
via the large global Ising-PPI-networks network with 8682 nodes.

The reduced Google matrix $\GR$ and its three components
$\Gpr$, $\Grr$, $\Gqr$ are shown in Figure~\ref{fig2}
for proteins of Table~\ref{table1} ($1 \leq i \leq 40$).
The weight of the component $\Gpr$ is $\Wpr=0.761$ 
being not so far from unity 
but this value is below $\Wpr \approx 0.95$
appearing usually in Wikipedia networks
\cite{regomax}; \cite{lirgomax}.
We attribute this to a significantly smaller number of links 
per node being $\ell = N_\ell/N \approx 3.1$
for the Ising-PPI-network
while for the English Wikipedia network of 2017 we have
$\ell \approx 22.5$ \cite{lirgomax}.
Indeed, the weight $\Wrr = 0.220$ 
of direct transitions of $\Grr$ is significantly larger than the 
corresponding values for the Wikipedia case with $\Wrr \approx 0.04$.
However, the weights $\Wqr = 0.019$ are comparable for both 
reduced networks. 

The matrix structure of direct transitions $\Grr$  has a clear two block
structure with dominant transitions inside each block 
associated to EGFR and JAK2 with only 4 significant matrix
elements from the EGFR to the JAK2 block. These matrix elements 
correspond to links from EGFR ($\pm$) to JAK2 (+) and STAT1(+) 
and have the same value $g\approx 0.0167$ 
while all other matrix elements (of this EGFR to JAK2 block) are very small
with the value $g \approx 1.73 \times 10^{-5}$ 
corresponding to the minimal value $(1-\alpha)/N$ in $G$ related to the 
damping factor $\alpha=0.85$. 

The matrix $\Gpr$ (which is exactly of rank 1) has a very simple structure 
with all columns being (approximately) proportional to the (local) PageRank 
of $\GR$ (which is itself proportional to the global PageRank projected 
onto the subset of 40 nodes) and one clearly sees that the strong 
horizontal red lines correspond 
to index positions $i$ of Table~\ref{table1} where the corresponding 
index $K$ is quite low below $\sim 100$ corresponding to a relatively high 
PageRank position. The full reduced matrix $\GR$ is numerically dominated by 
$\Gpr$ (but less clearly as for typical Wikipedia cases) and has at 
first sight a similar structure as $\Gpr$ but with somewhat smaller 
values. However, some of the strongest direct links (from $\Grr$) are also 
visible. 
Similarly to the Wikipedia network of politicians as discussed in 
\cite{regomax} both matrix components $\GR$ and $\Gpr$ are not very 
usefully to identify the indirect links. 

The indirect links are visible in the matrix $\Gqr$. As explained 
and shown mathematically 
in \cite{regomax} they correspond to pathways where a given 
node $i_1$ of the 
small subset points to a certain node outside the subset (in the big 
surrounding PPI network) which itself points eventually to another node 
outside the subset and comes after a finite number of iterations finally back 
to a different node $i_2$ inside the subset. This provides an indirect link 
from $i_1$ to $i_2$ and the weight or strength of this indirect link is 
characterized by the value of the matrix element 
$(\Gqr)_{i_2,i_1}$. According to Figure~\ref{fig2} there are now 
also significant interactions between the two blocks of EGFR and JAK2 
for the matrix $\Gqr$, sometimes with negative values (note that 
the matrix elements of $\Gqr$ may be negative). Figure~\ref{fig3} shows 
the the sum of the two components $\Grr+\Gqrnd$ ($\Gqrnd$ corresponds to 
$\Gqr$ without its diagonal elements) which confirms this observation. 
Actually, we consider that the elements of $\Grr+\Gqrnd$ describe best 
the combined direct and indirect links for the given subset. 

Due to the contribution of indirect transitions
there are additional transitions between these two blocks
where the four strongest additional elements of $\Gqr$ have values 
$g=0.0106$ ({\it GRB2 P62993 (+)} to {\it JAK2 O60674 (+)});
$g=0.0099$ ({\it GRB2 P62993 (+)} to {\it STAT1 P42224 (+)});
$g=  0.0059$ ({\it GAB1 Q13480 (+) } to {\it GTF2I P78347 (+)});
$g=  0.0039$ ({\it PIK3R1 P27986 (+) } to {\it GTF2I P78347 (+)}).
There are also 11 additional transitions
with $g > 0.1$. Thus even if the weight of $\Gqr$ 
is not high it provides important new indirect
interactions between proteins from the EGFR block to the JAK2 block.

\begin{figure}[h!]
\centerline{\includegraphics[width=0.48\textwidth]{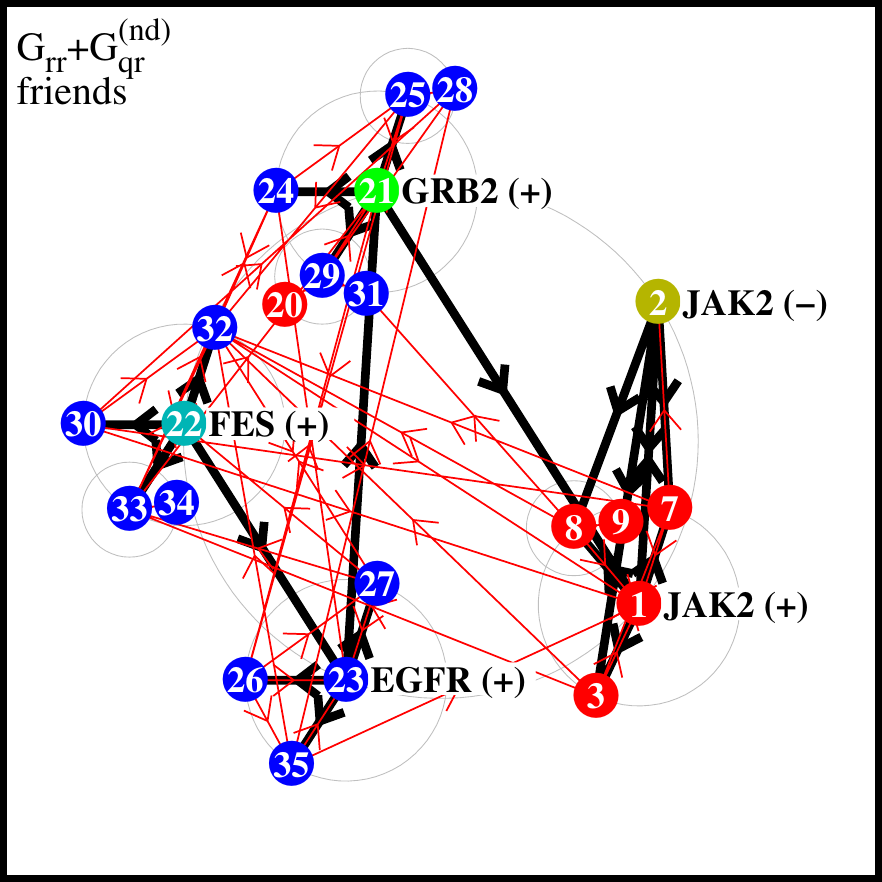}}
\caption{Network of friends for the subgroup of nodes 
given in Table~\ref{table1} 
corresponding to injection at {\it EGFR P00533 (+)} 
and absorption at {\it JAK2 O60674 (-)} constructed from the matrix 
$\Grr+\Gqrnd$ using 4 top (friends) links 
per column (see text for explanations). 
}
\label{fig4}
\end{figure}

The situation is even more striking when we consider
the transitions from the JAK2 block to the EGFR block. There are no direct 
links between these blocks in this direction from the global network 
but due to the construction 
of the Google matrix described above there are still numerically very small 
values $g$ for the matrix elements of $\Grr$ due to dangling nodes 
(nodes with no outgoing links) 
with $g=1/N\approx 1.15 \times 10^{-4}$ (in certain columns) 
or due to the damping factor term 
$(1-\alpha)/N\approx 1.73 \times 10^{-5}$ (for the other columns). 
On the other side concerning the indirect
links described by $\Gqr$ we find rather significant
transitions from the JAK2 block to the EGFR block with the four largest 
values: 
$g \approx 0.0122$ ({\it CCR2 P41597 (+)} to {\it  EGFR P00533 (-)} and to
{\it ESR1 P03372 (+)}); $g \approx 0.006 $ 
 ({\it CSF2RA/CSF2RB	SIGNOR-C212 (+)}  to
{\it ESR1	P03372 (+)} and to {\it PIK3R1	P27986 (+)}).
There are also 9 additional transitions with $g >0.001$.
Complete data files for the matrix elements of matrix components 
(for all examples) are available at \cite{ourwebpage}.

It is convenient to present 
the interactions between proteins, generated by
the matrix elements of the sum of 
two components $\Grr+\Gqrnd$ from Figure~\ref{fig3},
in the form of a network shown in Figure~\ref{fig4}.
To construct the network of effective friends,
we select first five initial nodes which are placed 
on a (large) circle: the two nodes with 
injection and absorption ({\it EGFR (+)} (injection node, blue)
and {\it JAK2 (-)} (absorption node, olive)
and three other nodes with a rather top position 
in the $K_L$ ranking: 
{\it JAK2 (+)} (related to {\it JAK2 (-)} with $K_L=1$, $i=1$, red), 
{\it GRB2 (+)} (with $K_L=2$, $i=21$, green) and 
{\it FES (+)} (with $K_L=3$, $i=22$, cyan). 
For each of these five initial 
nodes we determine four friends 
by the criterion of largest matrix elements (in modulus) in the same 
column, i.e. corresponding to the four strongest links 
from the initial node to 
the potential friends. The friend nodes found in this way are added to the 
network and drawn on circles of medium size around their initial node 
(if they do not already belong 
to the initial set of 5 top nodes). The links from the initial nodes to their 
friends are drawn as thick black arrows. For each of the newly added nodes 
(level 1 friends) we continue to determine the 
four strongest friends (level 2 friends) which are drawn on small circles and 
added to the network (if there are not already present from a previous 
level). The corresponding links from level 1 friends to level 2 friends are 
drawn as thin red arrows. 

Each node is marked by the 
index $i$ from the first column of Table~\ref{table1}.
The colors of the nodes are essentially red for nodes with strong negative 
values of $P_1$ (corresponding to the index $i=1,\ldots,20$) and blue 
for nodes with strong positive values of $P_1$ (for $i=21,\ldots,40$). 
Only for three of the initial nodes we choose different colors which 
are olive for {\it JAK (-)}, green for {\it GRB2 (+)} and cyan for  
{\it FES (+)}. This procedure generates the directed friendship
network shown in Figure~\ref{fig4}.

The obtained network of Figure~\ref{fig4} has
a rather clear separation between the two blocks related to EGFR
(mainly blue nodes) and JAK2 (mainly red nodes).
There is only one link of first level (black arrow)
from the EGFR block (GRB2 (+)) to the JAK2 block
(JAK2 (+)), Of course, there are other strong
direct transitions from the EGFR block to the JAK2 block as described 
above, but these links are weaker than the 4 closest friends and therefore 
they do not appear in the network structure of Figure~\ref{fig4}.
However, we see that there are many links between the two 
blocks on the secondary level of red arrows.

The block of JAK2 (red nodes) is rather compact with 
only 6 nodes (one red node 
at $i=20$ is more linked to the EGFR block).
In contrast the EGFR block contains 15 (blue) nodes
showing that this group of proteins
is characterized by broader and more 
extensive interconnections.
We think that such a network presentation
provides a useful qualitative image of the effective interactions between 
the two groups of proteins.

Network figures, for this example and the other two examples discussed below, 
constructed in the same way using the other matrix components 
$\GR$, $\Grr$ or $\Gqr$ (instead of $\Grr+\Gqrnd$) 
or using strongest matrix elements in rows (instead of columns) 
to determine follower networks are available at \cite{ourwebpage}. 

\subsection{Magnetization of proteins of EGFR - JAK2 pathway}

In the Ising-PPI-network each protein is described by two
components which can be considered as spin up or down state.
The PageRank probability of a protein
is given by the sum of probabilities of its two components
with $P(j) = P_+(j) + P_-(j)$. It can be shown that due to the structure of
the matrix transitions given by the matrices 
$\sigma_+ , \sigma_- , \sigma_0$ 
the sum of probabilities $P(j)$ for a given protein $j$
is the same  as for the directed PPI network
without doubling (see  Appendix).
Thus the activation or inhibition
links in the Ising-PPI-network
of doubled size only redistribute PageRank probability
for a given protein between up and down components.
The physical meaning of these up and down component probabilities
$P_+$ and $P_-$ is qualitatively related to the 
fact that on average the PageRank probability $P$ of a node
is proportional to the number of ingoing links.
Thus $P_+$ is proportional to the number of ingoing
activation links and  $P_-$ is proportional to the number of ingoing
inhibition links. Thus we can characterize
each node by its normalized magnetization
$M(j)=(P_+(j)-P_-(j))/(P_+(j)+P_-(j))$.
By definition $-1 \leq M(j) \leq 1$.
Big positive values of $M$ mean that
this protein has mainly ingoing activation links
while big negative values mean that
this protein has mainly inhibition ingoing links.
In principle, we can also study the magnetization
of CheiRank probability of proteins
given by $ M^*(j)=(P^*_+(j)-P^*_-(j))/(P^*_+(j)+P^*_-(j))$
but we keep this for further investigations.
We note that $M(j)$ and $ M^*(j)$ represent the normalized 
values which are independent of the total probability
$P(j), P^*(j)$. Thus the magnetization of nodes
of the reduced Google matrix remains the same as in the global network.


\begin{table}
\begin{center}
{
\relsize{-1}
\caption{Group of 38 nodes of the single protein network obtained 
from the group of Table~\ref{table1} by removing the $(+)$ and $(-)$ 
attributes. $K$ ($K^*$) represent the local rank indices obtained from 
the PageRank (CheiRank) ordering of the single protein network. 
The index $i$ is the same as in Table~\ref{table1} where the two values 
$i=2$ and $i=26$ do not appear here since they correspond to the two nodes 
where both components $(+)$ and $(-)$ are present in Table~\ref{table1}.}
\label{table2}
\begin{tabular}{rrrl}
\hline
$K$ & $K^*$ & $i$ &Node name \Tstrut\Bstrut\\
\hline\Tstrut
1 & 34 & 24 & PIK3CD O00329\\
2 & 3 & 1 & JAK2 O60674\\
3 & 1 & 23 & EGFR P00533\\
4 & 2 & 32 & PLCG1 P19174\\
5 & 10 & 21 & GRB2 P62993\\
6 & 11 & 28 & PTK2 Q05397\\
7 & 8 & 34 & ESR1 P03372\\
8 & 5 & 7 & STAT1 P42224\\
9 & 7 & 33 & SHC1 P29353\\
10 & 13 & 8 & MAP3K5 Q99683\\
11 & 4 & 25 & CBL P22681\\
12 & 25 & 31 & PIK3R1 P27986\\
13 & 6 & 37 & ERBB2 P04626\\
14 & 21 & 22 & FES P07332\\
15 & 12 & 5 & APOA1 P02647\\
16 & 9 & 19 & EZH2 Q15910\\
17 & 27 & 4 & ARHGEF1 Q92888\\
18 & 30 & 29 & GAB1 Q13480\\
19 & 28 & 3 & IFNGR2/INFGR1 SIGNOR-C142\\
20 & 24 & 30 & BCR P11274\\
21 & 22 & 40 & CRK P46108\\
22 & 26 & 27 & EZR P15311\\
23 & 15 & 6 & CSF2RA/CSF2RB SIGNOR-C212\\
24 & 19 & 38 & ERBB3 P21860\\
25 & 33 & 36 & SHC3 Q92529\\
26 & 18 & 10 & CCR2 P41597\\
27 & 14 & 39 & NCK1 P16333\\
28 & 31 & 15 & ITGAL P20701\\
29 & 17 & 9 & STAT4 Q14765\\
30 & 23 & 14 & CSF2RA P15509\\
31 & 16 & 20 & GTF2I P78347\\
32 & 37 & 16 & CTLA4 P16410\\
33 & 36 & 13 & EPOR P19235\\
34 & 20 & 18 & ITGB2 P05107\\
35 & 38 & 17 & STAP2 Q9UGK3\\
36 & 32 & 11 & PRMT5 O14744\\
37 & 35 & 12 & STAM Q92783\\
\Bstrut 38 & 29 & 35 & VAV2 P52735\\
\hline
\end{tabular}
}
\end{center}
\end{table}

\begin{figure}[h!]
\centerline{\includegraphics[width=0.48\textwidth]{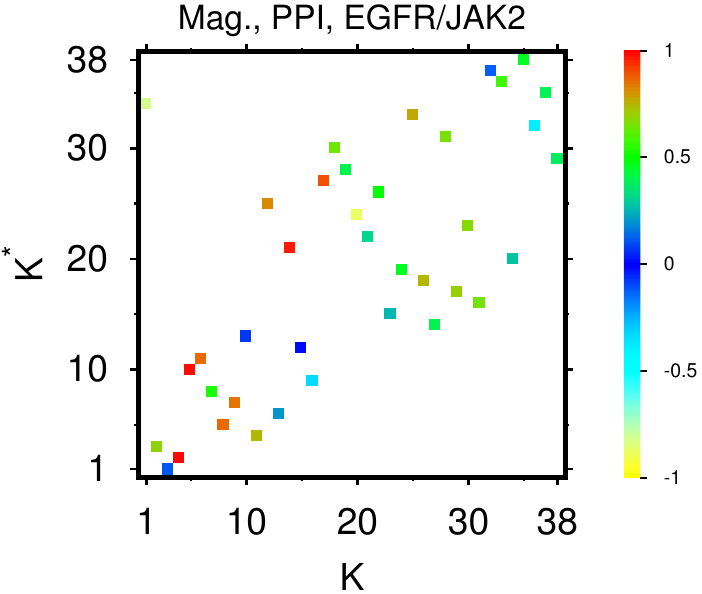}}
\caption{PageRank ``magnetization'' 
$M(j)=(P_+(j)-P_-(j))/(P_+(j)+P_-(j))$ of proteins of Table~\ref{table2}
shown on the PageRank-CheiRank plane $(K,K^*)$ of local indices;
here $j$ represents a protein node in the initial single protein network 
and $P_{\pm}(j)$ are the PageRank components of the Ising-PPI-network
(see text). 
The values of the color bar correspond to $M/\max{|M|}$ with 
$\max|M|=0.896$ being the maximal value of $|M(j)|$ for the shown group of 
proteins.}
\label{fig5}
\end{figure}

We take all different 38 proteins 
present in Table~\ref{table1} 
and consider their magnetization 
(this number
is smaller than 40 since for few 
proteins both $(+)$ or $(-)$ components are present in this Table).
All these 38 proteins are listed in Table~\ref{table2} with their local 
PageRank and CheiRank indices $K$ and $K^*$.
The distribution of these 38 proteins
on the PageRank-CheiRank plane is shown in Figure~\ref{fig5} and the colors of 
the square boxes presents the values of $M(j)$ (see caption of 
Figure~\ref{fig5}). 
The three proteins with the strongest positive magnetizations are
{\it PLCG1 P19174} ($M=0.8959$),
{\it GRB2 P62993} ($M=0.8899$),
{\it FES P07332} ($M=0.8719$)
and with the strongest negative values are
{\it BCR P11274} ($M= -0.7799$),
{\it PIK3CD O00329} ($M= -0.7328$),
{\it PRMT5 O14744} ($M=  -0.3527$).
In total there are only 5 proteins of Table~\ref{table2}
with negative magnetization values.
We attribute this to the fact that the number of inhibition
links is smaller than the number of activation ones.
We think that the magnetization of proteins can provide
new interesting information about the functionality of proteins.

\subsection{Examples of other protein  pathways}

We also consider two other proteins pairs for injection (pumping) and 
absorption which we analyzed in the same way. 
Again we compute the vector 
$V_0=G\,W_0$ where $W_0$ has only two non-zero components being $1$ 
at the pumping node and $-1$ at the absorption node, we solve the 
inhomogeneous PageRank equation (\ref{eq_PGinhom}) to obtain the linear 
response vector $P_1$ from which we determine a set of 40 nodes 
composed with 20 strongest negative and 20 strongest positive values. 
In order to ensure that the two initial injection and absorption nodes also 
belong to this subset we eventually replace the node at position 20 
for strongest positive and/or negative values with the injection and/or 
absorption node respectively. 
Here we only show and discuss the list of the obtained subsets and the 
effective network schemes corresponding to Table~\ref{table1} 
and Figure~\ref{fig4} for these two examples while Tables and Figures 
analogous to Table~\ref{table2}, Figures~\ref{fig1}, 
\ref{fig2}, \ref{fig3}, \ref{fig5} are given in Appendix.

First we discuss the case of injection at
{\it MAP2K1 Q02750 (+)} and absorption at {\it EGFR P00533 (-)}.
The protein MAP2K1 is a member of the dual-specificity protein kinase family 
that acts an integration point for multiple biochemical signals.
There is no direct link between  MAP2K1 and EGFR. 
The global PageRank indices of these two nodes are
$K= 84$ (PageRank probability $P(84)= 0.0009794$ ) for  
{\it MAP2K1 Q02750 (+)} and $K=136$ (PageRank probability $P(136)= 0.0007817$) 
for {\it EGFR P00533 (-)}. The subset of most sensitive proteins 
obtained from the LIRGOMAX algorithm for 
this protein pair is given in Table~\ref{table3}.  
These proteins are different from those of Table~\ref{table1}. 
We note now that the injection and absorption proteins
have lower positions in the rank indices $K_L$ and $i$ of Table~\ref{table3}. 
We attribute this somehow unexpected result of the $P_1$ ranking to 
rather nontrivial vortex flows on the Ising-PPI-network.

\begin{table}
\begin{center}
{
\relsize{-1}
\caption{Same as in Table~\ref{table1} 
but for injection (pumping) at {\it MAP2K1 Q02750 (+)}
and absorption at {\it EGFR P00533 (-)}.
}
\label{table3}
\begin{tabular}{rrrl}
\hline
$i$ & $K_L$ & $K$ &Node name \Tstrut\Bstrut\\
\hline\Tstrut
1 & 15 & 172 & FES	P07332 (+) \\
2 & 16 & 29 & GRB2	P62993 (+) \\
3 & 17 & 90 & EGFR	P00533 (+) \\
4 & 18 & 3 & PIK3CD	O00329 (-) \\
5 & 19 & 126 & CBL	P22681 (+) \\
6 & 20 & 648 & EZR	P15311 (+) \\
7 & 21 & 136 & EGFR	P00533 (-) \\
8 & 22 & 38 & PTK2	Q05397 (+) \\
9 & 23 & 30 & JAK2	O60674 (+) \\
10 & 24 & 57 & STAT1	P42224 (+) \\
11 & 26 & 456 & GAB1	Q13480 (+) \\
12 & 27 & 424 & BCR	P11274 (-) \\
13 & 29 & 9 & PI3K	SIGNOR-C156 (+) \\
14 & 32 & 26 & PLCG1	P19174 (+) \\
15 & 33 & 746 & SHC3	Q92529 (+) \\
16 & 34 & 2398 & VAV2	P52735 (+) \\
17 & 35 & 291 & ERBB2	P04626 (+) \\
18 & 36 & 15 & STAT3	P40763 (+) \\
19 & 37 & 888 & ERBB3	P21860 (+) \\
\Bstrut 20 & 38 & 40 & JAK1	P23458 (+) \\
\hline\Tstrut
21 & 1 & 125 & CEBPA	P49715 (+) \\
22 & 2 & 144 & MAPK14	Q16539 (-) \\
23 & 3 & 54 & GSK3B	P49841 (+) \\
24 & 4 & 543 & TAL1	P17542 (-) \\
25 & 5 & 74 & CASP9	P55211 (-) \\
26 & 6 & 16 & PPARG	P37231 (-) \\
27 & 7 & 1491 & ARRB2	P32121 (+) \\
28 & 8 & 156 & MAPK3	P27361 (+) \\
29 & 9 & 84 & MAP2K1	Q02750 (+) \\
30 & 10 & 246 & MAPK1	P28482 (+) \\
31 & 11 & 80 & IRS1	P35568 (-) \\
32 & 12 & 1 & CASP3	P42574 (+) \\
33 & 13 & 523 & KIF3A	Q9Y496 (+) \\
34 & 14 & 7 & ERK1/2	SIGNOR-PF1 (+) \\
35 & 25 & 528 & ERG	P11308 (+) \\
36 & 28 & 106 & MEF2C	Q06413 (+) \\
37 & 30 & 826 & ANGPT2	O15123 (+) \\
38 & 31 & 290 & TEK	Q02763 (+) \\
39 & 78 & 181 & CPT1B	Q92523 (+) \\
\Bstrut 40 & 86 & 20 & JUN	P05412 (+) \\
\hline
\end{tabular}
}
\end{center}
\end{table}

\begin{figure}[h!]
\centerline{\includegraphics[width=0.48\textwidth]{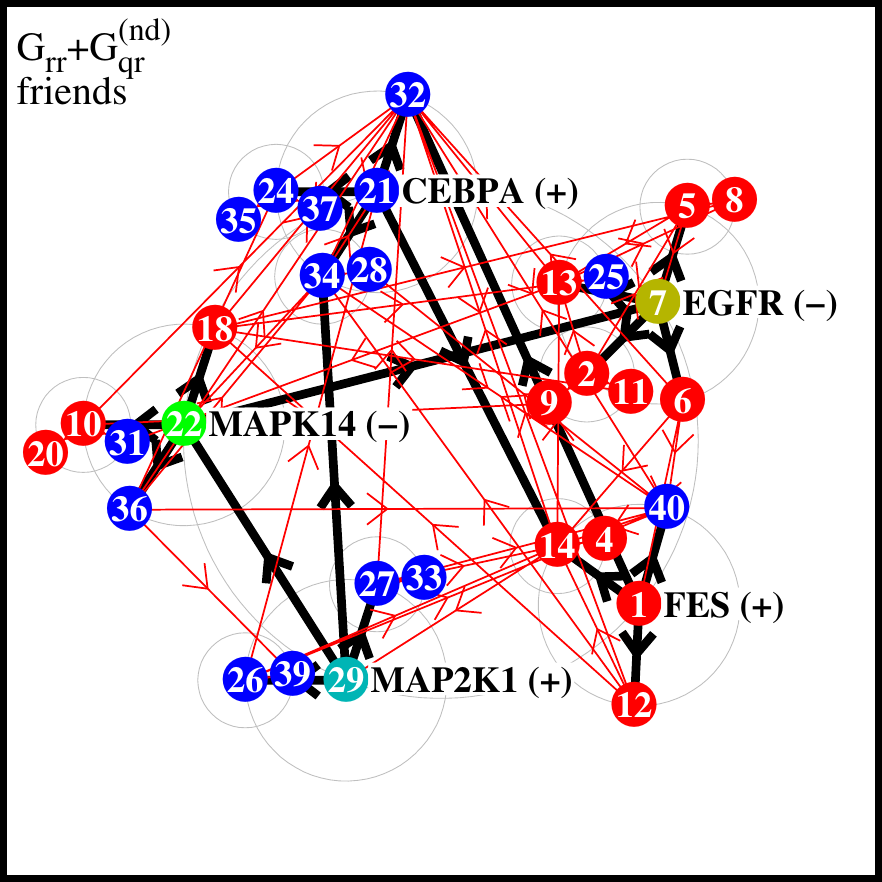}}
\caption{Same as Figure~\ref{fig4}
but for the pathway of Table~\ref{table3}.
}
\label{fig6}
\end{figure}

The friendship network for this case is shown in Figure~\ref{fig6} 
(the construction method is the same as Figure~\ref{fig4}). 
The 5 proteins of the initial large circle are
EGFR (-) (olive),
FES (+) (red),
MAP2K1 (+) (cyan),
MAPK14 (-) (green),
CEBRPA (+) (blue).
In this network we find a number of strong indirect links
from the block of {\it MAP2K1 Q02750 (+)} (blue nodes) to
{\it EGFR P00533 (-)} (red nodes) for which there is no direct link 
(e.g. from $i=21$ to $i=14$ proteins of Table~\ref{table3}).
In the opposite direction from red to blue nodes there are only
two strong direct matrix elements of $\Grr$ 
being from {\it PI3K SIGNOR-C156 (+)} $i=13$
to {\it IRS1 P35568 (-)}  $i=21$ with $g= 0.08501$
and from {\it STAT3 P40763 (+)} $i=18$ 
to {\it CASP3 P42574 (+)} $i=32$ with $g= 0.03543$
with all other elements being below $1.8 \times 10^{-5}$.
However, in this direction there are 9 new
indirect links with elements $g> 0.01$
and 20 with  $g> 0.005$.
This results in a rather dense network with many
links shown in Figure~\ref{fig6}.
From the network structure we see that the 
proteins $i=25, 40$ of the blue block
are more closely related with proteins of the 
red block and inversely the
proteins $i=10, 18, 20$ of the red block are more closely
related with proteins of the blue block.

As a further example we also briefly discuss the pathway
generated by injection at
{\it EGFR P00533 (+)} and absorption at
{\it PIK3CA P42336 (-)}.
These two proteins are conventional bio markers of lung cancer
(see e.g. \cite{krasnoyarsk}). 
The global PageRank indices of these two nodes are
$K= 90$ (PageRank probability $P(90)= 0.0009633$ ) for  {\it EGFR P00533 (+)} and
$K=1604$ (PageRank probability $P(1604)= 0.0001366$) for {\it PIK3CA P42336 (-)}.

The most sensitive proteins obtained by the LIRGOMAX algorithm,
are shown in Table~\ref{table4}. However, now 
the absorption node {\it PIK3CA P42336 (-)} 
has a very low value (in modulus) of 
$P_1$ ($P_1= -4.59 \times 10^{-5}$, $K_L= 2806$) 
and does initially not belong to the group of nodes with 20 top strongest 
negative values. Therefore we replace the node {\it AKT3 Q9Y243 (+)} ($K_L=70$) 
which was initially selected for $i=20$ by the absorption node 
{\it PIK3CA P42336 (-)}. 
Furthermore, also its $(+)$ component {\it PIK3CA P42336 (+)} 
($P_1= -0.004546$ and $K_L =138$) does not appear 
in Table~\ref{table4} showing that the influence of {\it EGFR P00533 (+)}
on the protein {\it PIK3CA P42336} is rather low.

\begin{table}
\begin{center}
{
\relsize{-1}
\caption{Same as in Table~\ref{table1} 
but for injection (pumping) at {\it EGFR P00533 (+)}
and absorption at {\it PIK3CA P42336 (-)}. 
}
\label{table4}
\begin{tabular}{rrrl}
\hline
$i$ & $K_L$ & $K$ &Node name \Tstrut\Bstrut\\
\hline\Tstrut
1 & 1 & 203 & BTK	Q06187 (+) \\
2 & 2 & 19 & AKT	SIGNOR-PF24 (+) \\
3 & 3 & 14 & AKT1	P31749 (+) \\
4 & 4 & 100 & AKT2	P31751 (+) \\
5 & 5 & 63 & MTOR	P42345 (+) \\
6 & 6 & 24 & PtsIns(3,4,5)P3	CID:24755492 (+) \\
7 & 7 & 80 & IRS1	P35568 (-) \\
8 & 8 & 23 & RAC1	P63000 (+) \\
9 & 10 & 330 & PI3K	SIGNOR-C156 (-) \\
10 & 11 & 9 & PI3K	SIGNOR-C156 (+) \\
11 & 38 & 1014 & TEC	P42680 (+) \\
12 & 39 & 970 & BMX	P51813 (+) \\
13 & 62 & 1587 & ITK	Q08881 (+) \\
14 & 63 & 154 & PIK3CB	P42338 (+) \\
15 & 65 & 1672 & DAPP1	Q9UN19 (+) \\
16 & 66 & 1076 & PLCG2	P16885 (+) \\
17 & 67 & 56 & mTORC1	SIGNOR-C3 (+) \\
18 & 68 & 1196 & GTF2I	P78347 (+) \\
19 & 69 & 75 & BAD	Q92934 (-) \\
\Bstrut 20 & 2806 & 1604 & PIK3CA	P42336 (-) \\
\hline\Tstrut
21 & 9 & 172 & FES	P07332 (+) \\
22 & 12 & 29 & GRB2	P62993 (+) \\
23 & 13 & 90 & EGFR	P00533 (+) \\
24 & 14 & 3 & PIK3CD	O00329 (-) \\
25 & 15 & 136 & EGFR	P00533 (-) \\
26 & 16 & 126 & CBL	P22681 (+) \\
27 & 17 & 30 & JAK2	O60674 (+) \\
28 & 18 & 57 & STAT1	P42224 (+) \\
29 & 19 & 648 & EZR	P15311 (+) \\
30 & 20 & 456 & GAB1	Q13480 (+) \\
31 & 21 & 424 & BCR	P11274 (-) \\
32 & 22 & 58 & SHC1	P29353 (+) \\
33 & 23 & 746 & SHC3	Q92529 (+) \\
34 & 24 & 2398 & VAV2	P52735 (+) \\
35 & 25 & 40 & JAK1	P23458 (+) \\
36 & 26 & 291 & ERBB2	P04626 (+) \\
37 & 27 & 888 & ERBB3	P21860 (+) \\
38 & 28 & 7 & ERK1/2	SIGNOR-PF1 (+) \\
39 & 29 & 1028 & JAK1/STAT1/STAT3	SIGNOR-C120 (+) \\
\Bstrut 40 & 30 & 303 & STAT1/STAT3	SIGNOR-C118 (+) \\
\hline
\end{tabular}
}
\end{center}
\end{table}

\begin{figure}[h!]
\centerline{\includegraphics[width=0.48\textwidth]{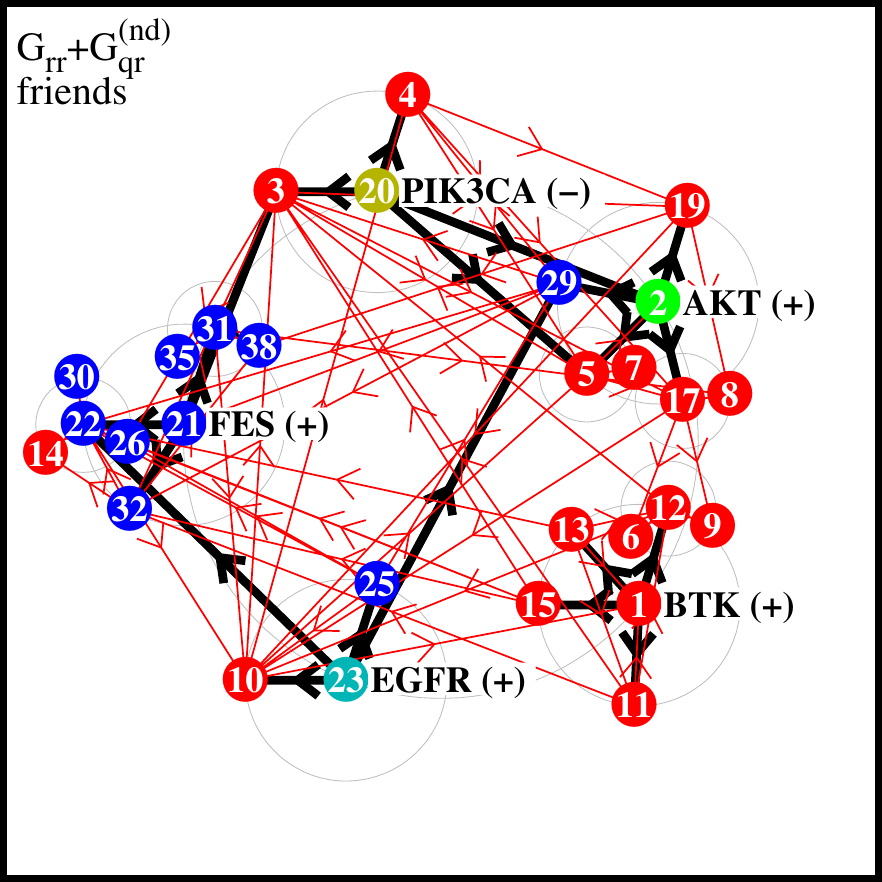}}
\caption{Same as Figure~\ref{fig4}
but for the pathway of Table~\ref{table4}.
}
\label{fig7}
\end{figure}

The friendship network structure of shown in Figure~\ref{fig7}
shows a clear separation between the two blocks
of positive (blue) and negative (red) $P_1$ values.
However, some proteins of one block
happen to be closer to proteins of the other block
(e.g. proteins $i=10, 14$ from the red block are closer to the blue block 
and blue block protein $i=29$ is closer to the proteins of the red block).
We also note that concerning the links from the blue to the red block there 
are 9 significant direct transitions (matrix elements of $\Grr$ larger 
than 0.01) and 35 significant indirect {\em and} direct transitions 
(matrix elements of $\Grr+\Gqrnd$ larger than 0.01). 
For the opposite direction of transitions from the red to the blue block
the increase is less significant but still there are new transitions
due to indirect pathways (2 significant transitions for $\Grr$
and 3 for $\Grr+\Gqr$).
The significance of indirect transitions
is also well visible in the friendship
network of Figure~\ref{fig7}
with many red arrows between the two blocks.

The same results for the original list, where the node {\it AKT3 Q9Y243 (+)}
at position $i=20$ has not been replaced by {\it PIK3CA P42336 (-)},
are available at \cite{ourwebpage}. 

\section{Discussion}

In this work we describe
the properties of Google matrix analysis of the 
bi-functional SIGNOR PPI network 
from \cite{signor}. The main elements of this approach are:
the activation and inhibition actions
of proteins on each-other are described by
Ising spin matrix transitions between 
the protein components in the doubled size Ising-PPI-network;
the recently developed LIRGOMAX \cite{lirgomax} algorithm 
determines the most sensitive proteins
on the pathway between two selected proteins
with probability injection (pumping)
at one protein and absorption at another protein;
the set of most sensitive proteins are analyzed by the 
REGOMAX algorithm which 
treats efficiently all direct and indirect
interactions in this subset 
taking into account all their effective interactions
through the global PPI network.
We illustrated the efficiency of this 
approach on several examples of two selected proteins.
The obtained  results show the  efficiency of 
the LIRGOMAX and REGOMAX algorithms.
We also show that the bi-functionality
of protein-protein interactions
leads to a certain effective 
magnetization of proteins
which characterizes their dominant action
on the global PPI network.

The executive codes and reduce Google matrix data
are open and publicly available at \cite{ourwebpage}
and interested researchers can easily
study any example of a pathway between any pair
of proteins from the SIGNOR network.

The described LIRGOMAX and REGOMAX algorithms can be applied also
to other type of biological networks
(e.g. metabolic networks discussed by \cite{metabol}).

We mention that the described Google matrix algorithms
have been tested for networks with 5 million nodes 
and thus they can operate efficiently 
on other PPI networks of significantly larger size
(e.g. MetaCore  network \cite{metacore}
which has several tens of thousands of nodes and about $2$ million links).  
Thus we expect that the Google matrix approach, or in short Googlomics,
will find broad applications for the analysis of protein-protein
interactions.

\section{Acknowledgments}
This work was supported in 
part by the Programme Investissements
d'Avenir ANR-11-IDEX-0002-02, 
reference ANR-10-LABX-0037-NEXT (project THETRACOM).
This work was granted access to the HPC resources of 
CALMIP (Toulouse) under the allocation 2019-P0110.

\onecolumn

\clearpage
{\parindent=0cm \bf \large APPENDIX \vspace*{0.3cm}}

\setcounter{equation}{0}
\setcounter{section}{0}
\setcounter{table}{0}
\renewcommand{\theequation}{A.\arabic{equation}}
\setcounter{figure}{0}
\renewcommand{\thesection}{A.\arabic{section}}
\renewcommand\thefigure{A.\arabic{figure}}
\renewcommand\thetable{A.\arabic{table}}

\section{Statistical properties of the SIGNOR protein-protein interactions
network PPI}

Using the Signor database  a network 
of $N=4341$ proteins with $N_\ell=12547$ interactions was created. 
In a first version, called the ``single protein network'', the links 
do not contain the information if the interaction corresponds to 
activation, inhibition or is neutral/unknown. As usual we first 
construct an adjacency matrix with entries $A_{ij}=1$ if there is a 
link from node $j\to i$ and $A_{ij}=0$ if there is no such link. However, 
in certain rare cases 
there are multiple types of links between two proteins (e.g. activation and 
inhibition) in which case we choose $A_{ij}$ being a multiplicity factor 
of $2$ or $3$ (instead of the usual entry $1$). 
Once the adjacency matrix is fixed the Google matrix of this (single) protein 
network is constructed in the usual way: 
column sum normalization, 
taking into account the effect of dangling nodes (nodes with no outgoing 
link) by replacing each zero column by a uniform column with entries $1/N$ 
and with the application of the standard damping factor $\alpha=0.85$. 

\begin{figure}[!h]
\centerline{\includegraphics[width=0.40\textwidth]{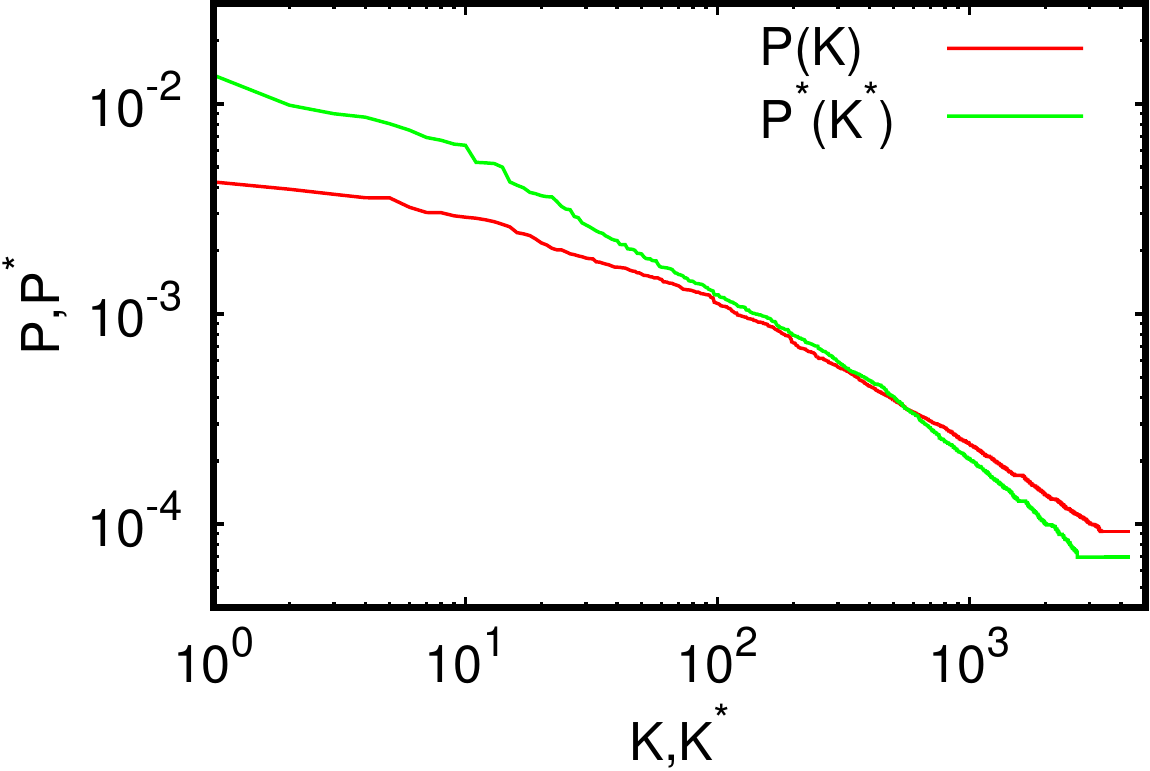}}
\caption{PageRank $P(K)$ and CheiRank $P^*(K^*)$ for the 
single protein network.}
\label{figS1}
\end{figure}

\begin{figure}[!h]
\centerline{\includegraphics[width=0.40\textwidth]{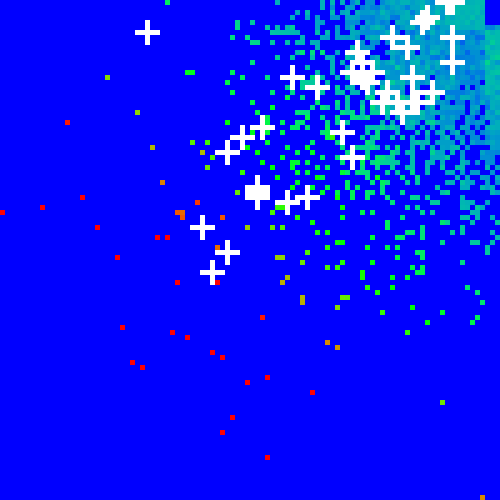}}
\caption{Density of nodes  $W(K,K^*)$ of the single protein network 
on PageRank-CheiRank plane $(K,K^*)$
averaged over $100\times100$ 
logarithmically equidistant grids for $0 \leq \ln K, \ln K^* \leq \ln N$,
the density is averaged over all nodes inside each cell of the grid,
the normalization condition is $\sum_{K,K^*}W(K,K^*)=1$.
The color bar of Fig.~2 applies (for positive values) and its values 
correspond to $(W/\max W)^{1/4}$. 
In order to increase the visibility large density values have 
been reduced to (saturated at) 1/16 of the actual maximum density.
The $x$-axis corresponds to $\ln K$ 
and the $y$-axis to $\ln K^*$. 
The white crosses show the positions of the 38 nodes of Table~2 
and in Fig.~5.}
\label{figS2}
\end{figure}

In Fig.~\ref{figS1} we show the PageRank $P$ (CheiRank $P^*$) 
for this single network 
versus the corresponding rank index $K$ ($K^*$) showing a typical decay 
(roughly) comparable to a power law $P\sim 1/K^\beta$ ($P^*\sim 1/(K^*)^\beta$)
with $\beta \approx 0.7$ ($0.8$) for $K\ge 100$ ($K^*\ge 10$). 
Fig. ~\ref{figS2} shows the density of nodes 
in the PageRank-CheiRank plane $(K,K^*)$ and the positions of the 
subgroup of nodes corresponding to Table~2 for this network. 

To take into account the information about the nature of the links we use 
the approach of the Ising-PageRank to construct a larger network 
where each node is doubled with two labels 
$(+)$ and $(-)$. To construct the doubled ``Ising'' network of proteins 
each unit entry of the initial adjacency matrix  is replaced by  
$2 \times 2$ matrices which are:
\begin{equation}
\label{eq_sigmadef}
\sigma_+=\left(\begin{array}{cc}
1 & 1 \\
0 & 0 \\
\end{array}\right)
\quad,\quad
\sigma_-=\left(\begin{array}{cc}
0 & 0 \\
1 & 1 \\
\end{array}\right)
\quad,\quad
\sigma_0=\frac12\,\left(\begin{array}{cc}
1 & 1 \\
1 & 1 \\
\end{array}\right)
\end{equation}
where $\sigma_+$ applies to ``activation'', $\sigma_-$ to ``inhibition'' 
and $\sigma_0$ to ``neutral'' or ``unknown''. For the rare cases 
with multiple types of links between two proteins 
we use the sum of the corresponding $\sigma$ matrices 
which increases the weight of the adjacency matrix elements. After this 
the corresponding Google matrix is constructed in the usual way. The doubled 
Ising protein network corresponds to $N_I=8682$ nodes and 
$N_{I,\ell}=27266$ links (according to the non-zero entries of the used 
$\sigma$ matrices).

Now the PageRank vector (of this doubles Ising network) 
has components $P_+(j)$ and $P_-(j)$. Due to the 
particular structure of the $\sigma$ matrices (\ref{eq_sigmadef}) one can 
show analytically  the exact identity $P(j)=P_+(j)+P_-(j)$ 
where $P(j)$ is the PageRank of the initial single protein network. For this 
we have to replace in Eq. (4) of \cite{isingnet} 
the value $n_i$ by 
$n_{ij}$ with $n_{ij}=1,0,1/2$ for the matrix $\sigma_+$, $\sigma_-$ or 
$\sigma_0$ respectively. The additional dependence of $n_{ij}$ on $j$ 
takes into account that the choice of the $\sigma$ matrix may be different 
for each link (and is not identical inside each row as it was the case for 
the model used in  \cite{isingnet}. Then the analytical 
argument of this work also applies in exactly the same way 
to the case of the doubled Ising protein network. 
We have also numerically verified that the identity 
$P(j)=P_+(j)+P_-(j)$ holds up to 
numerical precision ($\sim 10^{-13}$). 

\begin{figure}[!h]
\centerline{\includegraphics[width=0.40\textwidth]{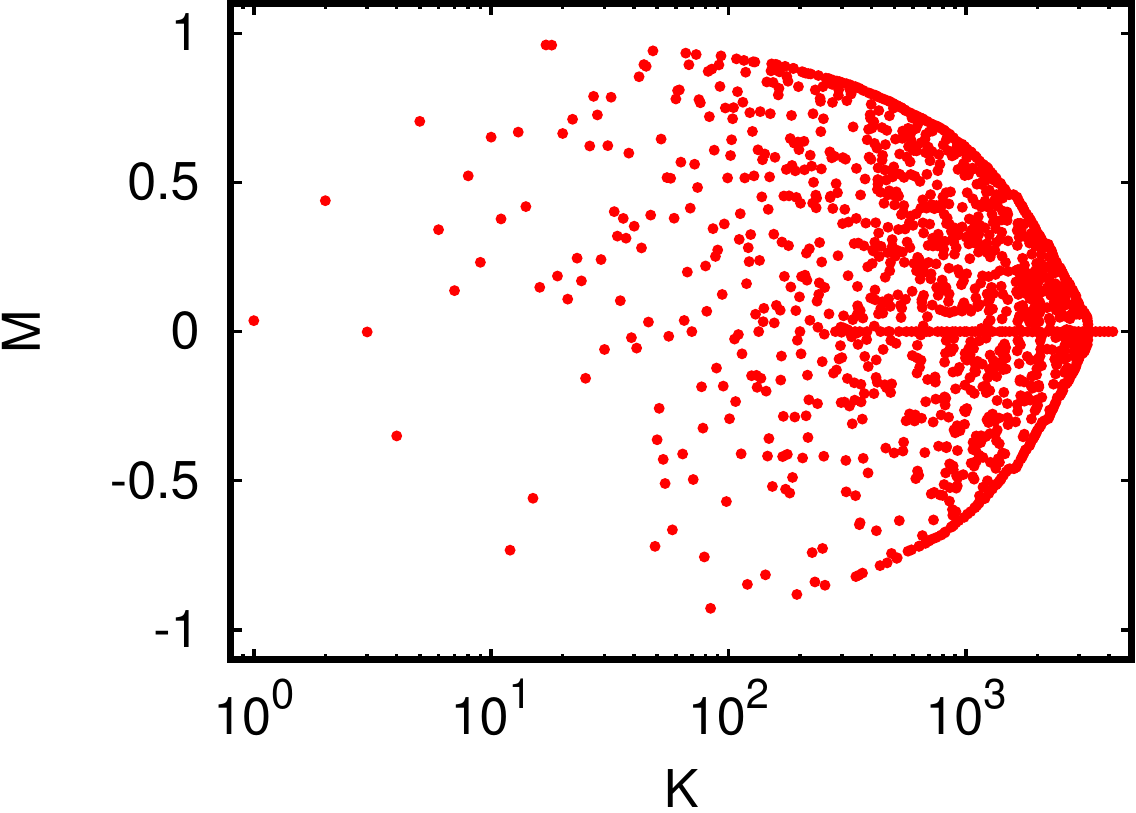}}
\caption{PageRank ``magnetization'' 
$M(j)=(P_+(j)-P_-(j))/(P_+(j)+P_-(j))$ in the Ising-PPI-network;
here $j$ is the node index and $K(j)$ is the PageRank index 
of the initial SIGNOR network
(without node doubling).}
\label{figS3}
\end{figure}

As in  \cite{isingnet} we introduce the 
PageRank ``magnetization''
by:
\begin{equation}
\label{eq_PRmagdef}
M(j)=\frac{P_+(j)-P_-(j)}{P_+(j)+P_-(j)}
\end{equation}
for a node $j$. 
The dependence of $M(j)$ on nodes is shown in Fig.~\ref{figS3} 
for the whole network and in Fig.~5 for the subgroup of 
nodes of Table~2. 

\newpage

\newpage


\section{Pathway from 
{\it MAP2K1 Q02750 (+) } to {\it EGFR P00533 (-)}}

Here we present additional figures and table for this pathway
discussed in subsection 4.3. Table~\ref{tableS1} gives the proteins 
(extracted from Table~3) for which the magnetization $M$ is presented 
in Figure~\ref{figS7}.

\begin{figure}[!h]
\centerline{\includegraphics[width=0.8\textwidth]{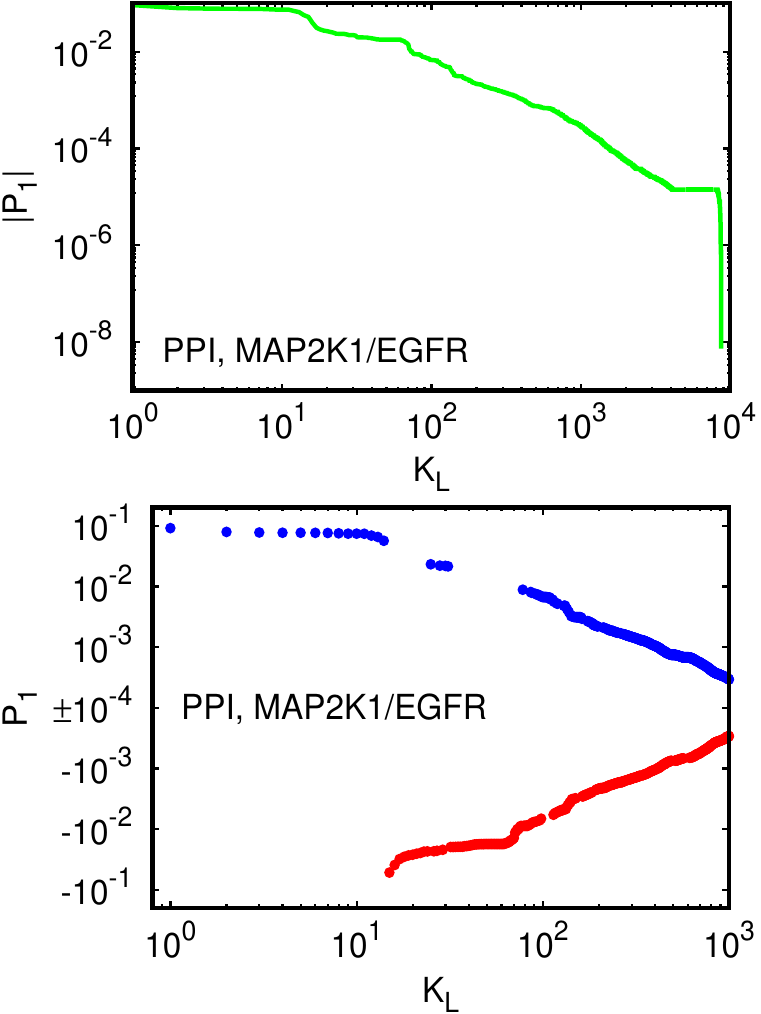}}
\caption{Same as in Fig.~1 but for the pathway
from {\it MAP2K1 Q02750 (+) } to {\it EGFR P00533 (-)}.}
\label{figS4}
\end{figure}

\begin{figure}[!h]
\centerline{\includegraphics[width=0.8\textwidth]{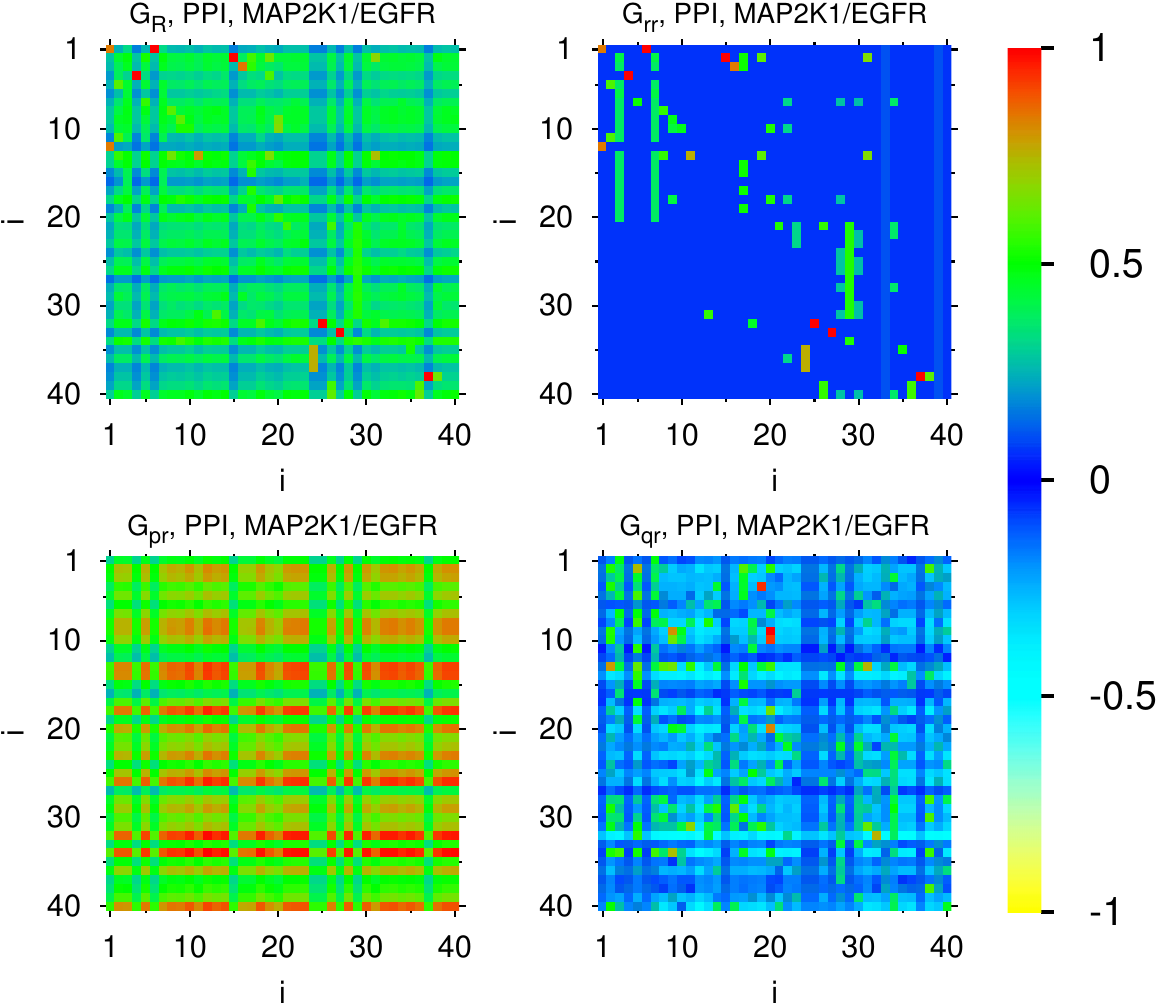}}
\caption{Same as in Fig.~2 but for the pathway
from {\it MAP2K1 Q02750 (+) } to {\it EGFR P00533 (-)}.}
\label{figS5}
\end{figure}

\begin{figure}[!h]
\centerline{\includegraphics[width=0.8\textwidth]{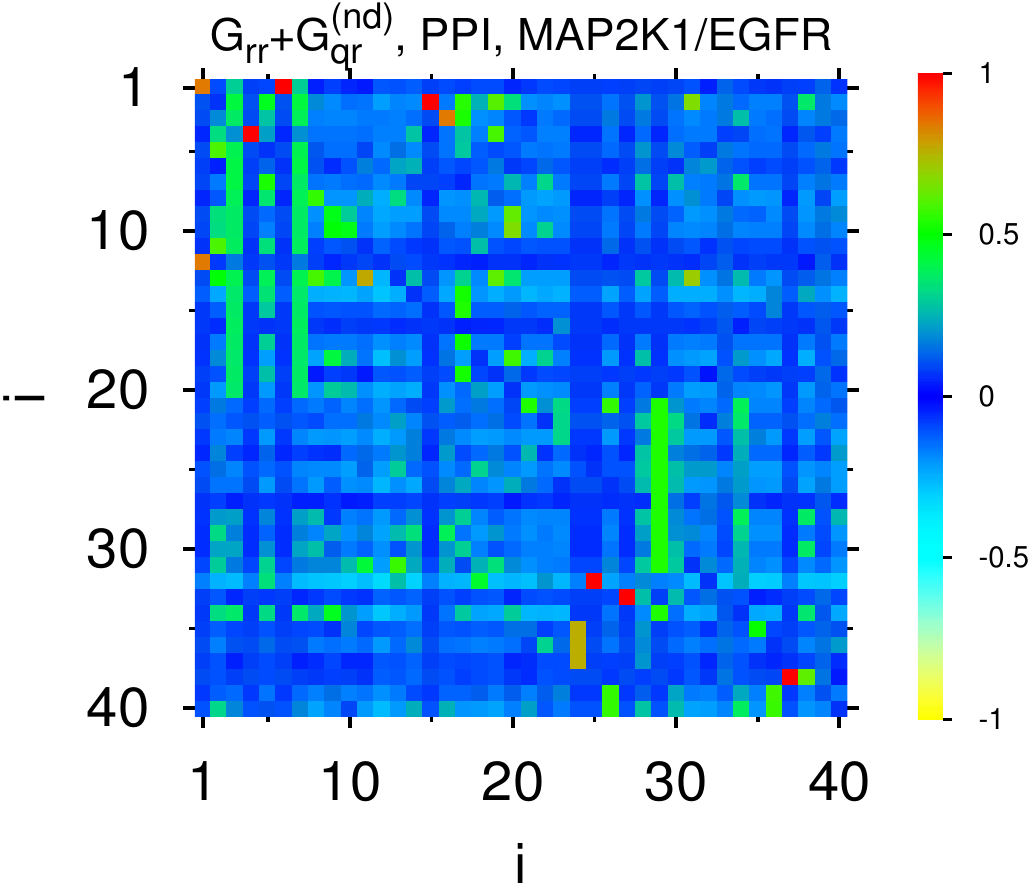}}
\caption{Same as in Fig.~3 but for the pathway
from {\it MAP2K1 Q02750 (+) } to {\it EGFR P00533 (-)}.}
\label{figS6}
\end{figure}
%
\twocolumn

\begin{figure}[!h]
\centerline{\includegraphics[width=0.7\textwidth]{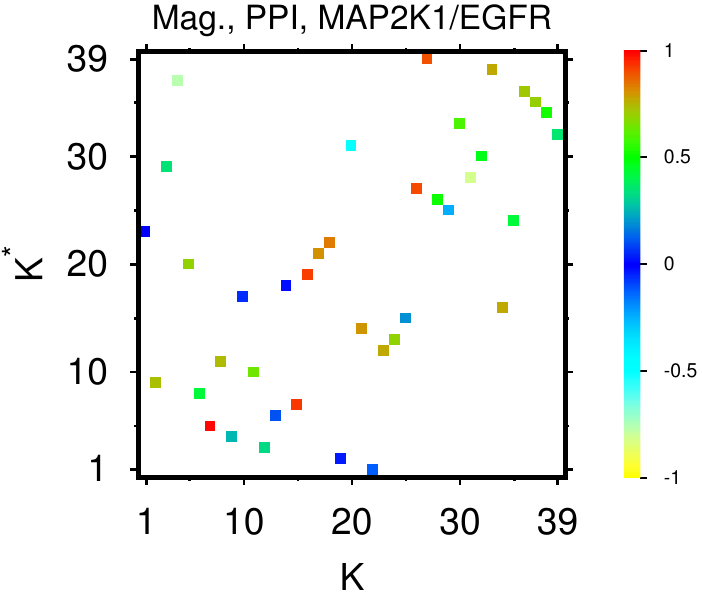}}
\caption{Same as in Fig.~5 but for the pathway
from {\it MAP2K1 Q02750 (+) } to {\it EGFR P00533 (-)}
with proteins from Table~\ref{tableS1}; 
the maximal magnetization used in the 
color bar normalization is $M_{max}=0.961$}
\label{figS7}
\end{figure}


\begin{minipage}[m]{.178\textwidth}
\phantom{aaaa aaa}
\end{minipage}
\begin{minipage}[m]{.41\textwidth}
\begin{center}
{
\relsize{-1}
\captionof{table}{Same as in Table~2 
but for injection (pumping) at {\it MAP2K1 Q02750 (+)}
and absorption at {\it EGFR P00533 (-)}. 
The index $i$ is the same as in 
Table~3 where two values do not appear here since they 
correspond to the two nodes where both components $(+)$ and $(-)$ are 
present in Table~3.
}
\label{tableS1}
\begin{tabular}{rrrl}
\hline
$K$ & $K^*$ & $i$ &Node name \Tstrut\Bstrut\\
\hline\Tstrut
1 & 23 & 26 & PPARG P37231\\
2 & 9 & 32 & CASP3 P42574\\
3 & 29 & 25 & CASP9 P55211\\
4 & 37 & 4 & PIK3CD O00329\\
5 & 20 & 13 & PI3K SIGNOR-C156\\
6 & 8 & 18 & STAT3 P40763\\
7 & 5 & 34 & ERK1/2 SIGNOR-PF1\\
8 & 11 & 40 & JUN P05412\\
9 & 4 & 22 & MAPK14 Q16539\\
10 & 17 & 36 & MEF2C Q06413\\
11 & 10 & 9 & JAK2 O60674\\
12 & 3 & 23 & GSK3B P49841\\
13 & 6 & 3 & EGFR P00533\\
14 & 18 & 21 & CEBPA P49715\\
15 & 7 & 14 & PLCG1 P19174\\
16 & 19 & 2 & GRB2 P62993\\
17 & 21 & 8 & PTK2 Q05397\\
18 & 22 & 20 & JAK1 P23458\\
19 & 2 & 28 & MAPK3 P27361\\
20 & 31 & 31 & IRS1 P35568\\
21 & 14 & 10 & STAT1 P42224\\
22 & 1 & 30 & MAPK1 P28482\\
23 & 12 & 29 & MAP2K1 Q02750\\
24 & 13 & 5 & CBL P22681\\
25 & 15 & 17 & ERBB2 P04626\\
26 & 27 & 1 & FES P07332\\
27 & 39 & 39 & CPT1B Q92523\\
28 & 26 & 38 & TEK Q02763\\
29 & 25 & 24 & TAL1 P17542\\
30 & 33 & 11 & GAB1 Q13480\\
31 & 28 & 12 & BCR P11274\\
32 & 30 & 6 & EZR P15311\\
33 & 38 & 33 & KIF3A Q9Y496\\
34 & 16 & 35 & ERG P11308\\
35 & 24 & 19 & ERBB3 P21860\\
36 & 36 & 15 & SHC3 Q92529\\
37 & 35 & 37 & ANGPT2 O15123\\
38 & 34 & 27 & ARRB2 P32121\\
\Bstrut 39 & 32 & 16 & VAV2 P52735\\
\hline
\end{tabular}
}
\end{center}
\end{minipage}


\onecolumn

\clearpage

\section{Pathway from 
{\it EGFR P00533 (+) } to {\it PIK3CA P42336 (-)}}

Here we present additional figures and table for this pathway
discussed in subsection 4.3. Table~\ref{tableS2} gives the proteins 
(extracted from Table~4) for which the magnetization $M$ is presented 
in Figure~\ref{figS11}.

\begin{figure}[!h]
\centerline{\includegraphics[width=0.8\textwidth]{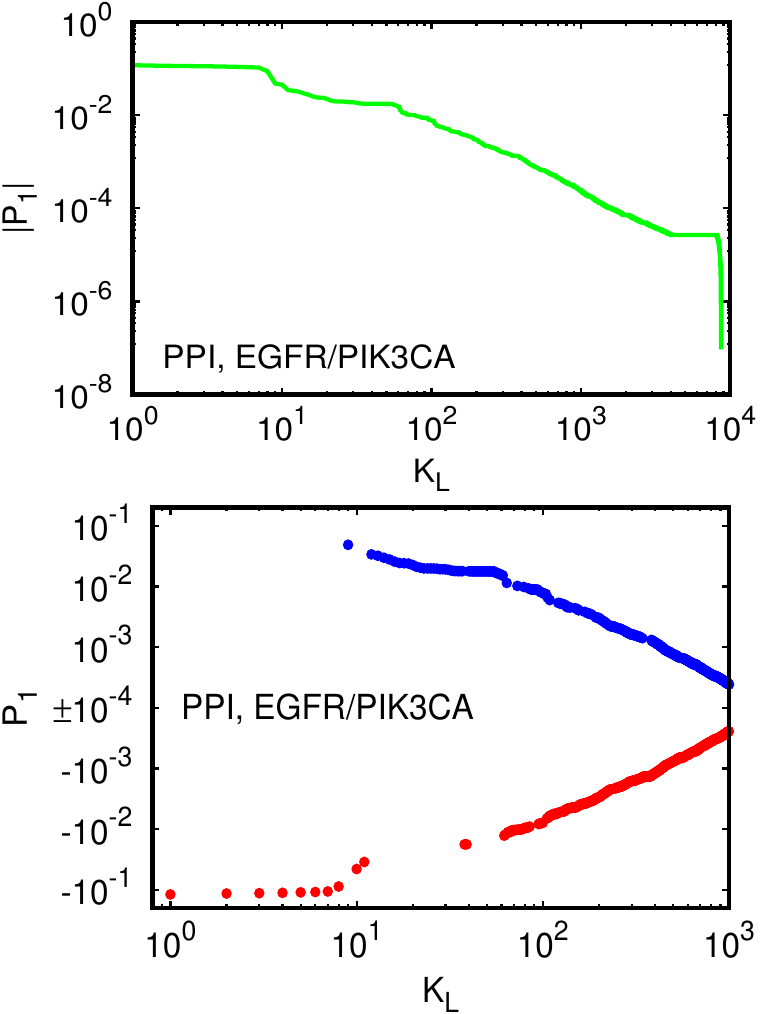}}
\caption{Same as in Fig.~1 but for the pathway
from {\it EGFR P00533 (+) } to {\it PIK3CA P42336 (-)}.}
\label{figS8}
\end{figure}

\begin{figure}[!h]
\centerline{\includegraphics[width=0.8\textwidth]{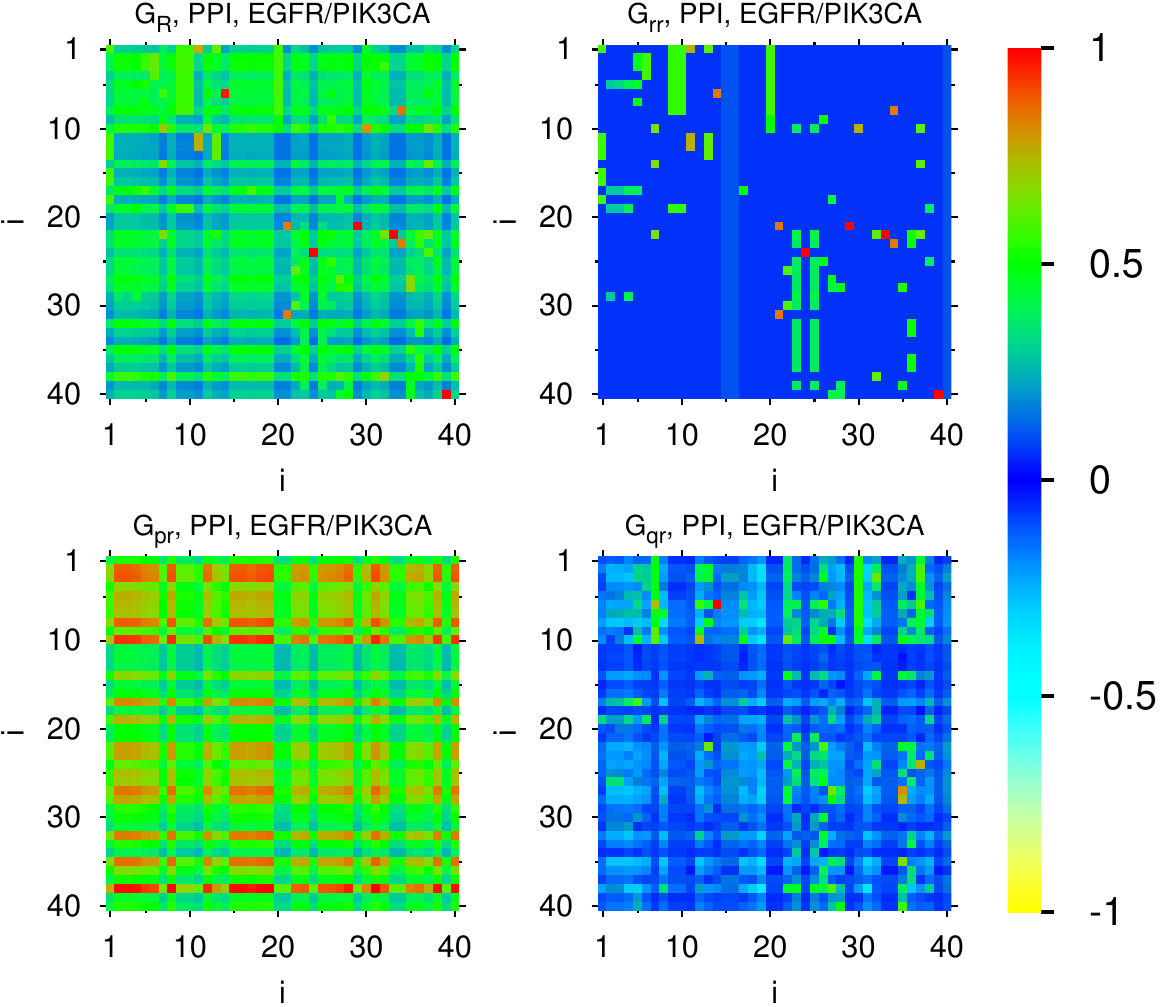}}
\caption{Same as in Fig.~2 but for the pathway
from {\it EGFR P00533 (+) } to {\it PIK3CA P42336 (-)}.}
\label{figS9}
\end{figure}

\begin{figure}[!h]
\centerline{\includegraphics[width=0.8\textwidth]{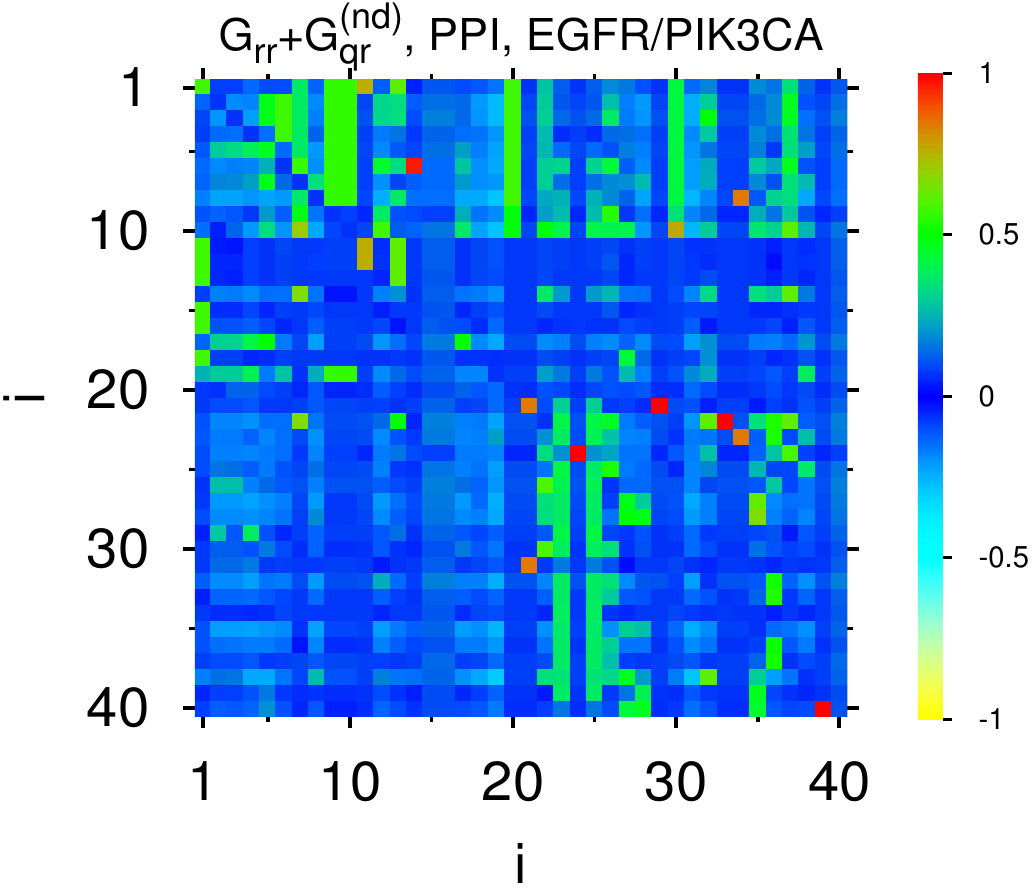}}
\caption{Same as in Fig.~3 but for the pathway
from {\it EGFR P00533 (+) } to {\it PIK3CA P42336 (-)}.}
\label{figS10}
\end{figure}

\twocolumn
\newpage

\begin{figure}[!h]
\centerline{\includegraphics[width=0.7\textwidth]{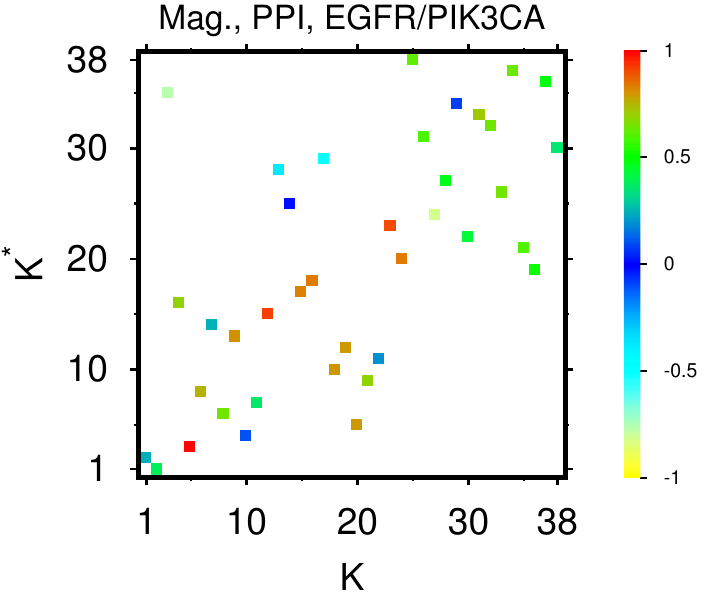}}
\caption{Same as in Fig.~5 but for the pathway
from {\it EGFR P00533 (+) } to {\it PIK3CA P42336 (-)}
with proteins from Table~\ref{tableS2};
the maximal magnetization used in the 
color bar normalization is $M_{max}=0.961$}
\label{figS11}
\end{figure}

%
\begin{minipage}[m]{.178\textwidth}
\phantom{aaaa aaa}
\end{minipage}
\begin{minipage}[m]{.41\textwidth}
\begin{center}
{
\relsize{-1}
\captionof{table}{Same as in Table~2 
but for injection (pumping) at {\it EGFR P00533 (+) } 
and absorption at {\it PIK3CA P42336 (-)}. 
The index $i$ is the same as in 
Table~4 where two values do not appear here since they 
correspond to the two nodes where both components $(+)$ and $(-)$ are 
present in Table~4.
}
\label{tableS2}
\begin{tabular}{rrrl}
\hline
$K$ & $K^*$ & $i$ &Node name \Tstrut\Bstrut\\
\hline\Tstrut
1 & 2 & 2 & AKT SIGNOR-PF24\\
2 & 1 & 3 & AKT1 P31749\\
3 & 35 & 24 & PIK3CD O00329\\
4 & 16 & 9 & PI3K SIGNOR-C156\\
5 & 3 & 38 & ERK1/2 SIGNOR-PF1\\
6 & 8 & 6 & PtsIns(3,4,5)P3 CID:24755492\\
7 & 14 & 17 & mTORC1 SIGNOR-C3\\
8 & 6 & 27 & JAK2 O60674\\
9 & 13 & 8 & RAC1 P63000\\
10 & 4 & 23 & EGFR P00533\\
11 & 7 & 5 & MTOR P42345\\
12 & 15 & 22 & GRB2 P62993\\
13 & 28 & 19 & BAD Q92934\\
14 & 25 & 14 & PIK3CB P42338\\
15 & 17 & 20 & PIK3CA P42336\\
16 & 18 & 35 & JAK1 P23458\\
17 & 29 & 7 & IRS1 P35568\\
18 & 10 & 28 & STAT1 P42224\\
19 & 12 & 32 & SHC1 P29353\\
20 & 5 & 4 & AKT2 P31751\\
21 & 9 & 26 & CBL P22681\\
22 & 11 & 36 & ERBB2 P04626\\
23 & 23 & 21 & FES P07332\\
24 & 20 & 1 & BTK Q06187\\
25 & 38 & 40 & STAT1/STAT3 SIGNOR-C118\\
26 & 31 & 30 & GAB1 Q13480\\
27 & 24 & 31 & BCR P11274\\
28 & 27 & 29 & EZR P15311\\
29 & 34 & 39 & JAK1/STAT1/STAT3 SIGNOR-C120\\
30 & 22 & 37 & ERBB3 P21860\\
31 & 33 & 33 & SHC3 Q92529\\
32 & 32 & 12 & BMX P51813\\
33 & 26 & 11 & TEC P42680\\
34 & 37 & 16 & PLCG2 P16885\\
35 & 21 & 18 & GTF2I P78347\\
36 & 19 & 13 & ITK Q08881\\
37 & 36 & 15 & DAPP1 Q9UN19\\
\Bstrut 38 & 30 & 34 & VAV2 P52735\\
\hline
\end{tabular}
}
\end{center}
\end{minipage}


\end{document}